\documentclass[preprint,12pt,authoryear]{elsarticle}
\RequirePackage{xcolor}

\usepackage{amssymb}
\usepackage{amsmath} 

\journal{Neuroimage}

\usepackage[margin=0.5in]{geometry}

\begin{document}

\begin{frontmatter}



\title{Multi-scale detection of hierarchical community architecture in structural and functional brain networks}


\author[a,b]{Arian Ashourvan}
\author[a,b]{Qawi K. Telesford}
\author[c]{Timothy Verstynen}
\author[b,a,d]{Jean M. Vettel}
\author[a,e,f]{Danielle S. Bassett}

\address[a]{Department of Bioengineering, University of Pennsylvania, Philadelphia, PA, 19104 USA}
\address[b]{U.S. Army Research Laboratory, Aberdeen Proving Ground, MD 21005 USA}
\address[c]{Department of Psychology, Center for the Neural Basis of Cognition, Carnegie Mellon University, Pittsburgh, Pennsylvania 15213}
\address[d]{Department of Psychological \& Brain Sciences, University of California, Santa Barbara,CA, 93106 USA}
\address[e]{Department of Electrical and Systems Engineering, University of Pennsylvania, Philadelphia, PA 19104 USA}
\address[f]{To whom correspondence should be addressed: dsb@seas.upenn.edu}

\begin{abstract}
Community detection algorithms have been widely used to study the organization of complex systems like the brain, which can be represented as graphs or networks of nodes (brain regions) connected by edges (functional or structural connections).  A principal appeal of these techniques is their ability to identify a partition of brain regions (or nodes) into clusters (or communities), where nodes within a community are densely interconnected. In their simplest application, community detection algorithms are agnostic to the presence of community hierarchies, but a common characteristic of many neural systems is a nested hierarchy with clusters embedded within clusters of other clusters. To address this limitation, we exercise a multi-scale extension of a common community detection technique known as modularity maximization, and we apply the tool to both synthetic graphs and graphs derived from human neuroimaging data, including structural and functional imaging data. Our multi-scale community detection algorithm links a graph to copies of itself across neighboring topological scales, thereby becoming sensitive to conserved community organization across neighboring levels of the hierarchy. We demonstrate that this method allows for a better characterization of topological inhomogeneities of the graph's hierarchy by providing a local (node) measure of community stability and inter-scale reliability across topological scales. We compare the brain's structural and functional network architectures and demonstrate that structural graphs display a wider range of topological scales than functional graphs. Finally, we build an explicitly multimodal multiplex graph that combines both structural and functional connectivity in a single model, and we identify the topological scales where resting state functional connectivity and underlying structural connectivity show similar \emph{versus} unique hierarchical community architecture. Together, our results showcase the advantages of the multi-scale community detection algorithm in studying hierarchical community structure in brain graphs, and they illustrate its utility in modeling multimodal neuroimaging data.\\
~\\
\end{abstract}


\end{frontmatter}



\newpage
\section{Introduction}

Hierarchical organization is a common motif in information processing systems \citep{bassett2010efficient}. The local embedding of similar processing units within groups that are then iteratively combined into larger and larger subsystems \citep{simon1991architecture} provides a unique solution to the problem of balancing information segregation (within a group at a single scale) and integration (between groups across multiple scales) \citep{park2013structural}. Such an organization is observed in very large-scale computer circuits and computing architectures \citep{ozaktas1992paradigms,chen2016vlsi}, cellular communication systems \citep{akyildiz2005wireless}, and social messaging systems \citep{moody2003structural}. Across these various real-world information processing systems, hierarchical organization can additionally offer robustness to damage \citep{zhang2007self,helbing2006information}, and a complex and diverse repertoire of system functions \citep{hilgetag2014hierarchical, valverde2015structural} that promote optimal and efficient information processing \citep{kinouchi2006optimal,beggs2008criticality} and transmission. 

While these previous examples are all man-made systems, hierarchical organization is also present in natural information processing systems. A quintessential example is the brain -- whether dissected in a nematode worm such as \emph{C. elegans}, or non-invasively measured in a healthy adult human \citep{lohse2014resolving}. Importantly, hierarchies in these systems can occur in both time \citep{chaudhuri2014diversity,siebenhuhner2013intra} and space \citep{bassett2010efficient}, and can exist in the clustering of gene expression \citep{barabasi2004network,conaco2012functionalization,arcila2014novel,henzler2013staged} or the groupings of neuronal cell types in lamina and columns \citep{sumbul2014genetic}. Arguably one of the most complex types of architecture in the brain is hierarchical \emph{network} architecture \citep{bassett2013multi,betzel2016multi}. Here, brain regions serve as nodes and structural or functional connections serve as edges in the network. Both structural and functional networks in the brain are critical conduits for information flow, processing, transmission, and cognitive computations more generally  \citep{sporns2010networks}. Importantly, hierarchical network structures can give rise to critical dynamics \citep{werner2009fractals}, where the behavioral repertoire of the neural system can be maximized with very few degrees of freedom \citep{chialvo2008brain}. Yet, despite its fundamental importance, our understanding of the hierarchical organization in the brain remains limited, in part due to the fundamental nature of complex networks: they defy visual interpretation, and instead require computational algorithms to  characterize. 

Algorithmic methods to identify hierarchical network architecture must overcome the challenge of identifying embedded processing units within local groups. Particularly useful candidates include community detection methods, which currently dominate the study of brain networks \citep{rubinov2010complex}. Community detection techniques can take on many forms \citep{porter2009,fortunato2010}, but perhaps the most common in the context of neuroimaging data is modularity maximization \citep{PhysRevE.69.026113}. In this approach, nodes are partitioned into communities such that nodes within a community are more likely to connect to one another than expected in a random network null model \citep{newman2010networks}. Importantly, the size of communities identified can be tuned by a structural resolution parameter, which titrates the relative difference between the real intra-community density and that expected in the null model \citep{reichardt2006statistical,porter2009communities}. Therefore, sweeping across a range of resolution parameters offers glimpses into the hierarchical organization of the graph \citep{fenn2009dynamic,fenn2012dynamical}; however, since the communities are identified independently at each point along the sweep, a secondary algorithm is required to track or link communities across topological scales, for example based on the similarity between communities in neighboring slices.

To address this limitation, we use a multi-scale community detection algorithm recently developed in applied mathematics \citep{mucha2010} to retrieve the underlying hierarchical organization of both artificial graphs and graphs representing human brain connectivity. We find that multi-scale community detection carefully preserves local information about the stability of sub-communities in the graph, enabling a thorough description of its hierarchical levels. Perhaps even more interestingly, we can uncover communities that remain stable across topological scales, and we can characterize their longevity and frequency. This approach offers unique advantages -- such as sensitivity to community longevity -- that extend more common methods that sweep across global topological scales \citep{fenn2009dynamic,pons2005computing} with independent estimates. Finally, we present methods for statistical assessment of the identified hierarchical communities, and we further offer an approach for the estimation of a consensus partition across the hierarchy.  

We exercise and apply this multi-scale approach to better understand the putative hierarchical community organization of patterns of white matter pathways (SC) estimated from diffusion spectrum imaging, and of functional connections (FC) estimated from resting state functional magnetic resonance imaging. Across 60 healthy adult individuals, we show that SC is topologically heterogeneous, displaying a varying number of stable communities across brain regions. In contrast, we show that the hierarchical organization of FC is flatter, displaying a smaller number of stable communities across scales. Building on these observations, we probe the spatial embedding of communities in each modality separately, and then compare and contrast the modalities with one another. Our work offers a roadmap for the use of multi-scale community detection in revealing hierarchical network structure in structural and functional brain graphs, and in assessing their relationships to one another. In future research, this technique could be combined with multilayer approaches to better understand multimodal hierarchical architectures in health and disease.

\section{Methods}

\subsection{Participants}
Sixty participants (28 male, 32 female) were recruited locally from the Pittsburgh, Pennsylvania area as well as the U.S. Army Research Laboratory in Aberdeen, Maryland. Participants were neurologically healthy adults with no history of head trauma, neurological or psychological pathology. Participant ages ranged from 18 to 45 years old (mean age, 26.5 years). Informed consent, approved by the Institutional Review Board at Carnegie Mellon University and in compliance with the Declaration of Helsinki, was obtained in writing for all participants. Pittsburgh participants were financially compensated for their time.
 
\subsection{MRI acquisition}

All 60 participants were scanned at the Scientific Imaging and Brain Research Center at Carnegie Mellon University on a Siemens Verio 3T magnet fitted with a 32-channel head coil. An MPRAGE sequence was used to acquire a high-resolution (1 mm$^{3}$ isotropic voxels, 176 slices) T1-weighted brain image for all participants. DSI data was acquired following fMRI sequences using a 50 min, 257-direction, twice-refocused spin-echo EPI sequence with multiple $q$ values ($TR =11,400 ms$, $TE =128 ms$, voxel size 2.4 mm$^{3}$, field of view 231 $\times$ 231 mm, $b$-max 5000 s/mm$^{2}$, 51 slices). Resting state fMRI (rsfMRI) data consisting of 210 T2*-weighted volumes were collected for each participant with a BOLD contrast with echo planar imaging (EPI) sequence (TR 2000 ms, TE 29 ms, voxel size 3.5 mm$^{3}$, field of view 224 $\times$ 224 mm, flip angle $79$ degrees).
 
Head motion is a major source of artifact in resting state fMRI data (rsfMRI). Although recently developed motion correction algorithms are far more effective than typical procedures \citep{satterthwaite2013improved,power2014methods,prium2015evaluation,rastko2016benchmarking}, head motion was additionally minimized during image acquisition with a custom foam padding setup designed to minimize the variance of head motion along pitch and yaw directions. The setup also included a chin restraint that held the participant's head to the receiving coil itself. Preliminary inspection of EPI images at the imaging center showed that the setup minimized resting head motion to 1 mm maximum deviation for most subjects. Only 2 subjects were excluded from the final analysis because they moved more than 2 voxels multiple times throughout the imaging session.

\subsection{Diffusion MRI reconstruction}

DSI Studio (http://dsi-studio.labsolver.org) was used to process all DSI images using a $q$-space diffeomorphic reconstruction method \citep{yeh2011ntu}. A nonlinear spatial normalization approach \citep{ashburner1999nonlinear} was implemented through 16 iterations to obtain the spatial mapping function of quantitative anisotropy (QA) values from individual subject diffusion space to
the FMRIB 1 mm fractional anisotropy (FA) atlas template. QA is an orientation distribution function (ODF) based index that is scaled with spin density information that permits the removal of isotropic diffusion components from the ODF to filter false peaks, facilitating the resolution of fiber tracts using deterministic fiber tracking algorithms. For a detailed description and comparison of QA with standard FA techniques, see \cite{yeh2013deterministic}. The ODFs were reconstructed to a spatial resolution of 2 mm$^{3}$ with a diffusion sampling length ratio of 1.25. Whole-brain ODF maps of all 60 subjects were averaged to generate a template image of the average tractography space. 

\subsection{Fiber tractography and analysis}

Fiber tractography was performed using an ODF-streamline version of the FACT algorithm \citep{yeh2013deterministic} in DSI Studio (September 23, 2013 and August 29, 2014 builds). All fiber tractography was initiated from seed positions with random locations within the whole-brain seed mask with random initial fiber orientations. Using a step size of 1 mm, the directional estimates of fiber progression within each voxel were weighted by $80\%$ of the incoming fiber direction and $20\%$ of the previous moving direction. A streamline was terminated when the QA index fell below $0.05$ or had a turning angle greater than $75$ degrees. We performed a region-based tractography to isolate streamlines between pairs of regional masks. All cortical masks were selected from an upsampled version of the original Automated Anatomical Labeling Atlas (AAL) \citep{tzourio2002automated,desikan2006automated} containing 90 cortical and subcortical regions of interest but not containing cerebellar structures or the brainstem. This resampled version contains 600 regions and is created via a series of upsampling steps in which any given region is bisected perpendicular to its principal spatial axis in order to create 2 equally sized sub-regions \citep{hermundstad2013structural,hermundstad2014structurally}. The final atlas contained regions of an average size of 268 voxels (with a standard deviation of 35 voxels). Diffusion-based tractography has been shown to exhibit a strong medial bias \citep{croxson2005quantitative} due to partial volume effects and poor resolution of complex fiber crossings \citep{jones2010twenty}. To counter the bias away from more lateral cortical regions, tractography was generated for each cortical surface mask separately. 

\subsection{Resting state fMRI preprocessing and analyses}

SPM8 (Wellcome Department of Imaging Neuroscience, London) was used to preprocess all rsfMRI collected from 53 of the 60 participants with DSI data. To estimate the normalization transformation for each EPI image, the mean EPI image was first selected as a source image and weighted by its mean across all volumes. Then, an MNI-space EPI template supplied with SPM was selected
as the target image for normalization. The source image smoothing kernel was set to a FWHM of 4 mm, and all other estimation options were kept at the SPM8 defaults to generate a transformation matrix that was applied to each volume of the individual source images for further analyses. The time-series was up-sampled to a 1Hz TR using a cubic-spline interpolation. Regions from the AAL600 atlas were used as seed points for the functional connectivity analysis \citep{hermundstad2013structural,hermundstad2014structurally}. A series of custom MATLAB functions were used to do the following: (1) extract the voxel time series of activity for each region, (2) remove estimated noise from the time series by selecting the first five principle components from the white matter and CSF masks.

\subsection{Data preprocessing}

Both the DSI and BOLD data were used to construct $(N\times N,  N=600 ~ regions)$ structural and functional networks. We then studied the hierarchical community structure of these graphs using multi-scale community detection. 

\subsubsection{Functional network construction}

Following prior work \citep{bassett2011,mantzaris2013dynamics,bassett2013task,bassett2014cross}, we estimated the dynamic functional connectivity between all region pairs using a wavelet coherence \citep{grinsted2004}. We choose the wavelet decomposition based on its denoising properties \citep{zhang2016choosing} and its utility in estimating statistical similarities between long memory time series such as those observed in resting state fMRI data \citep{achard2008fractal}. We observe two distinct bands of high coherence: $0.24-0.17Hz$ and $0.16 - 0.08 Hz$. We focus on the $0.16 - 0.08Hz$ band due to known sensitivity to underlying neural activity \citep{hutchison2013dynamic}. Coherence amplitudes were averaged over all frequencies and time points within the selected band to construct the average band-passed coherence for each pair of regions resulting in a single $N \times N$ functional connectivity (FC) adjacency matrix per subject \citep{chai2017evolution}.

\subsubsection{Structural network construction}

The individual subject's structural connectivity (SC) matrix represents the fiber count between all region pairs. Commonly the fiber count values are normalized by the region size, so that the values of the SC matrix reflect the \emph{density} of the white matter streamlines constructed between two regions \citep{hagmann2008mapping,bassett2011conserved,gu2015controllability,betzel2016optimally}. However, we circumvented this issue by using the AAL600 atlas \citep{hermundstad2013structural,hermundstad2014structurally}, which was purposefully designed to contain similarly-sized regions. Due to the heavy tailed nature of the edge weight distribution (see Appendix Fig.~\ref{fig:SI_Fig_edgedist}) we applied a log transform to the edge weights $\textstyle(log(SC+1)/max(log(SC+1)))$ (see Fig.~\ref{fig:ResFigure1}A) to increase the discriminibility of low edge weights and well as to increase the comparability to the edge weight distribution of functional connections.

\subsection{Community detection}

Common community detection algorithms can be used to partition a graph into clusters, where nodes tend to be more tightly connected to other nodes in their same cluster than to nodes in other clusters \citep{porter2009,fortunato2010}. In the context of network data (or other relational data that can be represented as a network), we adopt common parlance and refer to these clusters as \emph{communities}. A common heuristic for the identification of community structure in network data is the optimization of a quality function, which measures the relative density of the intra- \emph{versus} inter-community edges \citep{PhysRevE.69.026113,reichardt2004detecting}. One particularly popular quality function is the \emph{modularity} quality function \citep{newman2006modularity}, which can be defined as: 
 
\begin{equation}
Q = \sum_{ij} \left [   \left ( A_{ij}-\gamma P_{ij} \right )\right ]\delta \left ( g_{i},g_{j} \right )~\\,
\end{equation}

\noindent where for a graph of $N$ nodes, $A$ is the $N \times N$ weighted adjacency matrix, the $ij^{th}$ element of the adjacency matrix indicates the weight of the connection between node $i$ and node $j$, the Kronecker delta $\delta \left ( g_{i},g_{j} \right) =1 $ if the community assignment of node $i$ and node $j$ ($g_{i},g_{j}$) are identical ($g_{i} = g_{j}$) and zero otherwise, $\gamma$ is the structural resolution parameter, and $P_{ij}$ is the expected weight of the $ij^{th}$ edge between node $i$ and node $j$ under a specified null model. Importantly, by maximizing this quality function, one can identify a partition of nodes into communities; however, identifying the optimal partition is NP-hard, and therefore the problem is usually solved with clever heuristics such as the Louvain-like locally greedy algorithm \citep{blondel2008fast}. To account for the near degeneracy of the modularity landscape \citep{good2010performance}, the algorithm is used to optimize the modularity quality function multiple times, and results are only reported that remain consistent over those optimizations \citep{bassett2013robust}.

The Newman-Girvan null model \citep{girvan2002} is the most commonly used null model in modularity maximization. It can be defined as: $P_{ij} = \frac{k_{i}k_{j}}{2m}$, where  $k_{i} =\sum_{j} A_{ij}$ is the strength of node $i$ and $m = \frac{1}{2} \sum_{ij} A_{ij}$. In short, this null sets the expectations of an edge based on the strength of its nodes. The choice of structural resolution parameter $\gamma =1$ is common \citep{lancichinetti2011,berry2011tolerating,traag2011narrow}, however it only represents the community organization at a single topological scale \citep{bassett2013robust,lohse2014resolving}. Because graphs often display hierarchical organization, it is frequently useful to explore community structure in a graph over a range of values for $\gamma$ \citep{fenn2009dynamic,fenn2012dynamical,bassett2013robust,lohse2014resolving}. When a graph has a particularly salient topological scale at which community structure is strongest, this parameter sweep can be used to identify the ``optimal'' structural resolution parameter value at which this community structure can be identified \citep{bassett2013robust}. However, for a graph that has hierarchical structure in which multiple topological scales are equally salient, this approach can fail to identify a single ``optimal'' structural resolution parameter value.

\begin{figure*}
\centering
\includegraphics[width=.7\linewidth]{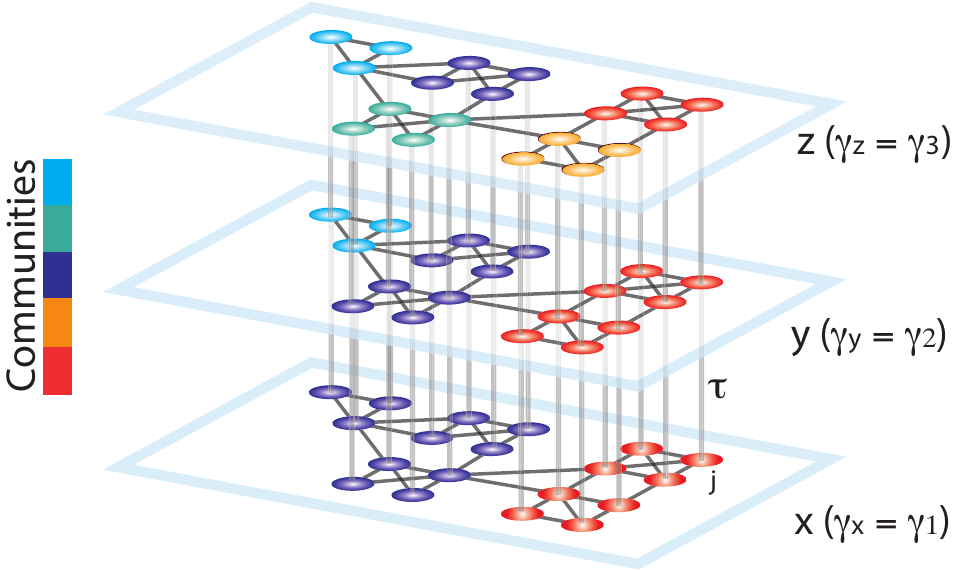}
\caption{\textbf{Schematic representing the construction of a multilayer network for use in multi-scale modularity maximization}. Duplicates of a graph are connected in a multilayer fashion to construct a 3D graph. The smallest resolution parameter $\gamma$ is assigned to the first layer ($x$), and it is linearly increased for the neighboring layers ($y,z$). The topological scale coupling parameter, $\tau$, tunes the strength of dependence of the communities across layers. Since the community assignments are dependent on the adjacent layers, nodes that display high clustering over neighboring topological scales are identified as a single community spanning several scales. In this schematic, the large communities identified at initial layers progressively break into smaller sub-communities, revealing the hierarchical community organization of the graph.}  
\label{fig:Figure1}
\end{figure*}

\subsection{Hierarchical community detection}

To study hierarchical community structure in graphs, we suggest a method based on optimizing the modularity quality function across all neighboring topological scales simultaneously. We achieve this by creating a multilayer network \citep{mucha2010}, linking duplicates of the graph at each $\gamma$ value to the graphs at neighboring $\gamma$ values. For a schematic representation of the proposed multilayer graph, see Fig.~\ref{fig:Figure1}. Formally, we define the multi-scale modularity quality function as:

\begin{equation}
Q = \frac{1}{2\mu }\sum_{ijxy} \left \{ \left ( A_{ij}-\gamma_{x} P_{ij} \right )\delta_{xy} +\delta_{ij}\tau_{jxy}  \right \}\delta \left ( g_{ix},g_{jy} \right )~\\,
\end{equation}

\noindent where the $ij^{th}$ element of the adjacency matrix $\mathbf{A}$ indicates the weight of the connection between node $i$ and node $j$, the Kronecker delta $\delta \left ( g_{ix},g_{jy} \right )=1$  if the community assignments of node $i$ from scale $x$ and node $j$ from scale $y$ ($g_{ix},g_{jy} $) are identical ($g_{ix}=g_{jy} $) and zero otherwise, $\gamma_{x}$ is the structural resolution parameter at layer $x$, $P_{ij}$ is the expected weight of the edge between node $i$ and node $j$, $\tau_{jxy}$ is the \emph{topological scale coupling parameter} which indicates the strength of the links between neighboring topological scales (as represented by layers), the total edge weight in the network is $\mu =\frac{1}{2}\sum_{jy}^{ } K_{jy}$, the strength of node $j$ in layer $y$ is $K_{jy} = k_{jy}+c_{jy}$, the intra-layer strength of node $j$ in layer $y$ is $k_{jy} =\sum_{i}^{ } A_{ij}$, and the inter-layer strength of node $j$ in layer $y$ is $c_{jy} =\sum_{x}^{ } \tau_{jxy}$. 

Following prior work \citep{wymbs2012differential,bassett2013robust,bassett2015extraction,papadopoulos2016evolution}, here we choose the expected values of the edge $P_{ij}$ uniformly for all edges as the average strength of all nodes: $P_{ij}$ equals some constant. This is sometimes referred to as the \emph{geographic} null model \citep{bassett2015extraction,papadopoulos2016evolution}. We choose the constant pragmatically, and separately for the structural matrices \emph{versus} functional matrices. In the structural matrices, we noted that the $min(A_{ij})$ is relatively consistent across the subjects in our sample, since the smallest value of streamline count is 1 streamline, while $mean(A_{ij})$ is quite different across subjects in our sample. By contrast, in functional matrices, we noted that the $mean(A_{ij})$ was relatively consistent across subjects in our sample, while the $min(A_{ij})$ was not. In order to maintain the greatest sensitivity to structure that is conserved across subjects in the sample, we therefore chose $P_{ij} =  min(A_{ij})$ for structural matrices and $P_{ij} =  mean(A_{ij})$ for functional matrices. Note that this means that the exact value of $\gamma$ used across structural and functional matrices is not directly comparable, while its relative value is. 

In this manuscript, we only examine the multi-scale community structure of the FC and SC graphs at a low value of the topological scale coupling parameter $(\tau = 0.5)$ where the community organization of the neighboring topological scales exhibit relatively small dependencies on one another. While not the focus of this paper, it is important to note that the hierarchical community detection framework we describe and exercise can also be extended to other multi-layer and temporal graphs. In the latter case, our framework can be used to link communities across different temporal scales; see Appendix.A for details.

 \begin{figure*}
\centering
\includegraphics[width=1\linewidth]{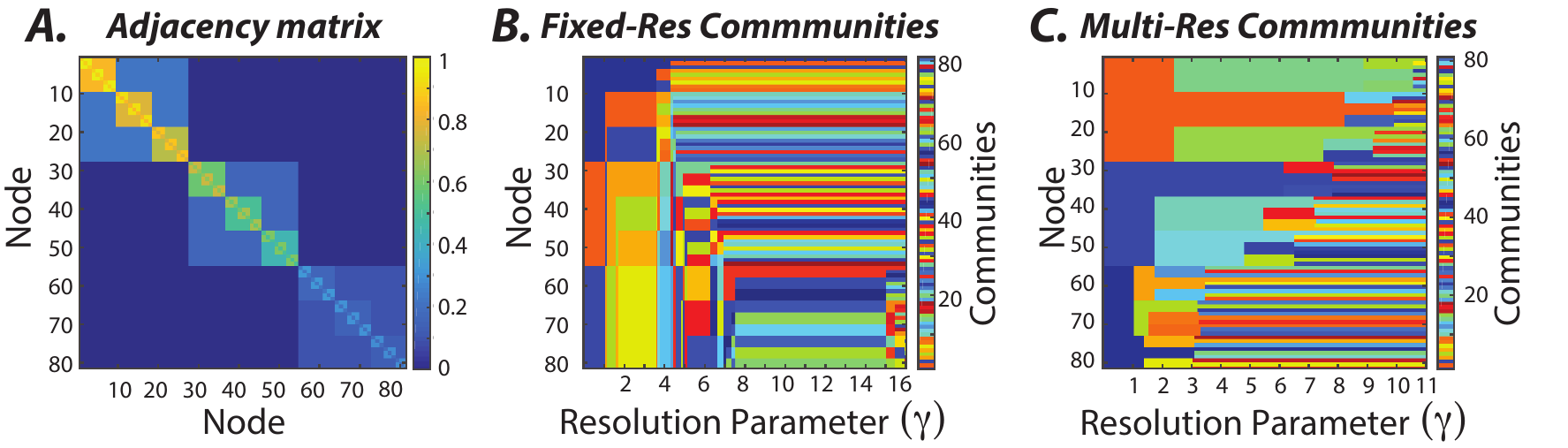}
\caption{\textbf{Uncovering hierarchical community structure in a synthetic graph.} \emph{(A)} Graphs can display heterogeneity in hierarchical community structure. To gain intuition regarding the utility of our method for characterizing these sorts of graphs, here we design a synthetic graph such that each node is part of a small cluster composed of 3 nodes, a medium sized cluster composed of 9 nodes, and a large cluster composed of 27 nodes. Heterogeneity is introduced by adding gradients in the values of edge weights such that not all clusters of a given size have the same average weight. \emph{(B)} We seek to uncover the hierarchical structure in this synthetic graph, first using the more traditional approach: maximizing a single-layer modularity quality function \citep{newman2006modularity} with the Newman-Girvan null model \citep{girvan2002} using a Louvain-like locally greedy algorithm \citep{blondel2008fast}. We sweep the resolution parameter between 1 and 16, and identify communities independently at each $\gamma$ value. The limitation of this approach is that there is no guaranteed correspondence between communities at one resolution and communities at another resolution. \emph{(C)} To overcome this limitation, we next seek to uncover the hierarchical structure in this synthetic graph using a multi-scale approach built on multi-scale community detection \citep{mucha2010}. We find that the hierarchical community detection uncovers the true underlying community organization over a continuous variation in the value of the structural resolution parameter. Moreover, the $\gamma$ value at which a community is detected tracks the mean edge weight of the community; stronger communities are identified at larger $\gamma$ values, and weaker communities are identified at smaller $\gamma$ values.}
\label{fig:Figure2}
\end{figure*}

\subsection{Statistics of multi-scale sommunities: Stability and consensus communities}

The multi-scale community detection algorithm identifies many communities that could span several topological scales, each here represented as a layer in the multilayer network. In other similar multilayer contexts, it is crucial to be able to assess the stability of the identified communities across scales \citep{fenn2009dynamic}, under the assumption that stable communities are of particular interest \citep{lambiotte2014random}. In our multi-scale framework, we measure the stability of individual nodes' allegiance to their communities across scales: for node \textit{i} the stability of its allegiance to community \emph{'X'} is calculated as the number of $\gamma$ values where node \textit{i} belongs to community \emph{'X'} divided by the total number of slices (i.e., all structural resolution parameter values examined). Higher values of stability indicate that a node belongs to a single community across a greater number of layers, indicating its participation in a wider range of topological scales in the hierarchy. 

Importantly, a node's stability can be calculated as a function of the value of the structural resolution parameter, $\gamma$. For example, suppose node \textit{i} is assigned to community \emph{'X'} at $\gamma=1$ and community \emph{'Y'} at $\gamma=2$. The stability of node \textit{i} at the point $\gamma=1$ is then equal to the number of $\gamma$ values where node \textit{i} belongs to community \emph{'X'} divided by the total number of slices; by contrast, the stability of the same node \textit{i} at the point $\gamma=2$ is equal to the number of $\gamma$ values where node \textit{i} belongs to community \emph{'Y'} divided by the total number of slices. Thus, in fact we can calculate a stability matrix that encodes the stability of each node at each value of the structural resolution parameter (e.g., Appendix Fig.~\ref{fig:ResFigure3}A). In this matrix, highly diverse patterns of stability are indicative of topological heterogeneity in the graph. By constrast, less diverse patterns of stability are indicative of topological homogeneity in the graph.

Because we employ a heuristic to maximize the modularity quality function \citep{blondel2008fast}, the identified partition of the multilayer network into multi-scale communities can change at each iteration \citep{good2010performance}. To establish a robust, representative partition across these iterations, we perform the following steps: (i) we maximize the modularity quality function many times $(n =100)$ to adequately sample the modularity landscape, (ii) for every pair of brain regions, we calculate the average probability of two nodes appearing in the same community (which we refer to as the \emph{intra-layer community allegiance}) from the multi-scale partitions for all layers, (iii) for every brain region, we calculate the average probability of it appearing in the same community across two neighboring layers (which we refer to as the \emph{inter-layer community allegiance}) from the multi-scale partitions for all neighboring layers, (iv) we identify the nodes (and layers) with reliable inter-and intra-layer community allegiance by comparing average community allegiance values with that of a null model. The average community allegiance of the null model was generated from randomizing community labels from step (i). Then, (v) we create a consensus multi-layer graph where the values of the intra-layer and intra-layer edges correspond to the average community allegiance of the edges that were found to be significantly different from the null model. All the non-significant edges were removed from the multi-layer graph. Finally, (vi) the consensus partition is identified from the multi-layer consensus graph using the multi-layer community detection algorithm with parameter values $\gamma =1$, and $\omega =1$.

\section{Results}

\subsection{Hierarchical community organization of synthetic graphs}

To illustrate the method, we begin with a synthetic hierarchical graph that is constructed so as to contain clear community structure across a range of topological scales (Fig.~\ref{fig:Figure2}A). Specifically, the graph displays identifiable community structure across four topological scales, with nested clusters of 3, 9, and 27 nodes.  Heterogeneity is introduced by adding gradients in the values of edge weights such that not all clusters of a given size have the same average weight. To uncover the hierarchical community structure in this synthetic graph, we first applied an existing approach: a maximization of a single-layer modularity quality function \citep{newman2006modularity} with the Newman-Girvan null model \citep{girvan2002} using a Louvain-like locally greedy algorithm \citep{blondel2008fast}. Across different values of the structural resolution parameter ($\gamma$), we observe that the communities identified appear to change frequently, with different communities being present at different values of $\gamma$ (Fig.~\ref{fig:Figure2}B). Prior work suggests that a reasonable method to choose the ``optimal'' value for the structural resolution parameter is to identify a range of $\gamma$ over which the community structure does not change appreciably \citep{fenn2009dynamic,bassett2013robust}. Applying that approach to these data, one might identify $9 < \gamma < 15$ as a range of $\gamma$ values over which the community structure is relatively stable (Fig.~\ref{fig:Figure2}B). Yet, the community structure present in this range of $\gamma$ values alone reveals little about the planted hierarchical organization and the topological inhomogeneities across nodes, as seen in Fig.~\ref{fig:Figure2}A.
 
To overcome this limitation, we apply a multi-scale community detection method using the hierarchical algorithm described in the Methods section. We observe that the community structure displays branching across layers (or values of $\gamma$; Fig.~\ref{fig:Figure2}C), meaning that a community in one layer can branch into two or more subcommunities in the next layer. Tracking the changes in a node's community allegiance as a result of branching into subcommunities gives us information about the local topology of the synthetic graph. These results suggest that the multi-scale community detection technique can accurately uncover planted hierarchical communities in synthetic graphs. To gain further intuition regarding the performance of the method, we also apply the approach to three other synthetic graphs with differing architectures (Appendix Fig.~\ref{fig:Figure3}). Again, we observe that the community structure displays branching across layers and that the number of branches is indicative of the number of local hierarchical scales in the synthetic graph. Moreover, across all synthetic graphs, we can observe that communities with higher-valued edge weights appear at higher values of $\gamma$ than communities with lower-valued edge weights. Together, these examples highlight the utility of the multi-scale community detection method for revealing hierarchically organized communities.

\subsection{Hierarchical community organization of white matter structure in the human brain}

\begin{figure*}
\centering
\includegraphics[width=1\linewidth]{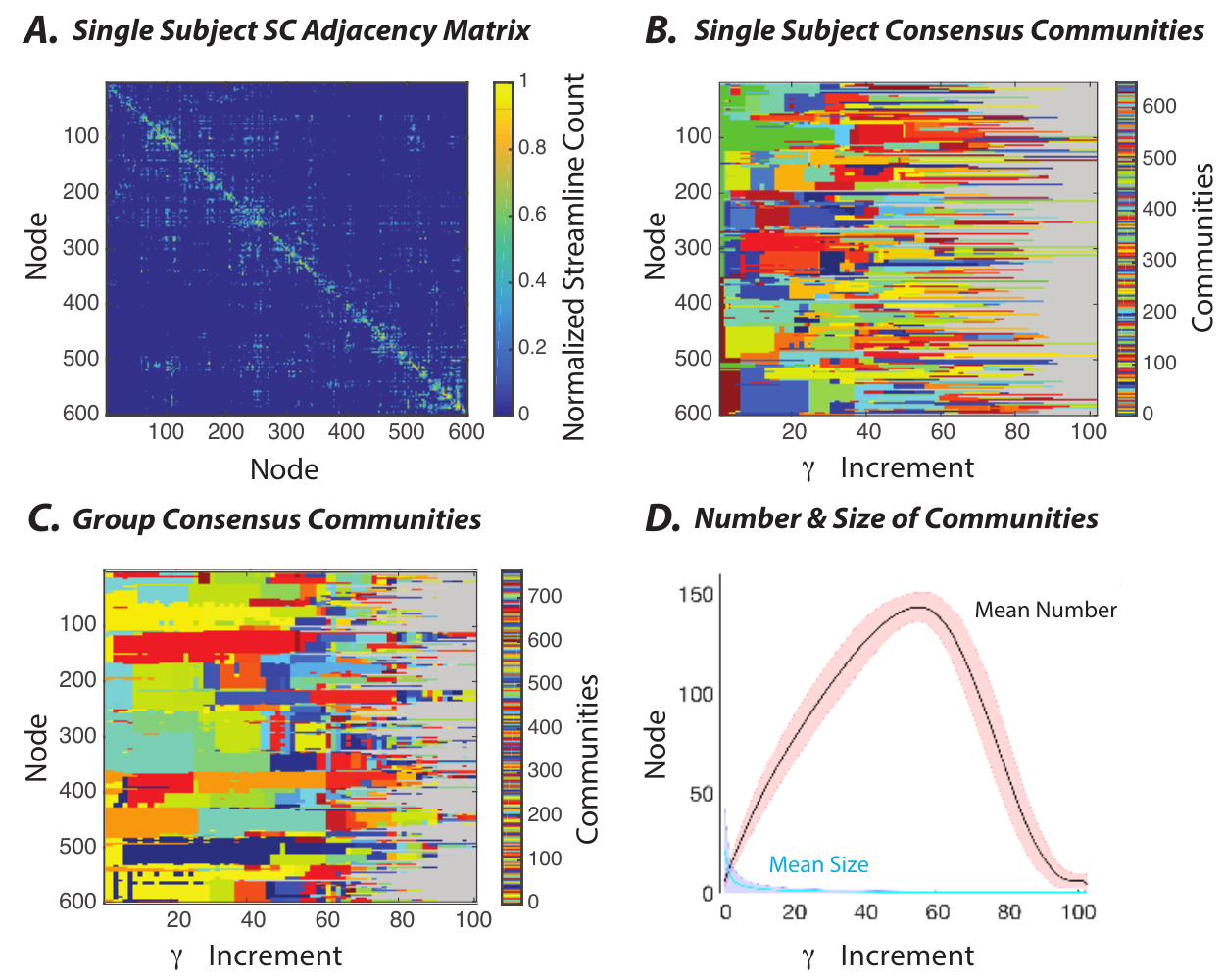}
\caption{\textbf{Application of multi-scale community detection to subject-level and group-level structural brain networks.} \emph{(A)} An example structural connectivity matrix from one subject, in which each element linking a pair of brain regions represents the number of streamlines reconstructed between those two areas. To better visualize its structure, we apply a log transformation $\textstyle(log(SC+1)/max(log(SC+1)))$. \emph{(B)} The consensus partition representing the multi-scale community structure for the matrix in panel \emph{(A)}. To enhance the visual detection of communities, we have represented all singleton communities with the same gray color. \emph{(C)} The group-level consensus partition representing the multi-scale community structure for the structural matrix, which is defined as the consensus over all participants' partitions. Here, again, to enhance visual clarity, we color the singleton communities in the same gray color. \emph{(D)} The average number as well as the average size (expressed as the percentage of total nodes) of the non-singleton communities calculated across layers (i.e., $\gamma$ increments). In these analyses, we used a $ \gamma \in [0.0133, 1] $, an inter-layer $\gamma$ increment $= 0.0133$, and 75 layers.} 
\label{fig:ResFigure1}
\end{figure*}

For a given subject, the structural connectivity (SC) matrix generated from the fiber count between 600 brain parcels is sparse, with an average streamline count of $3.57$ and a standard deviation of $37.83$ (see Methods). Intuitively, this sparsity indicates that relatively few brain regions share direct fiber connections with one another. Moreover, the distribution of edge weights is heavy tailed (see Fig.~\ref{fig:SI_Fig_edgedist}), ranging from only a few streamlines per node pair to several thousand per node pair. To better visualize the architecture of the SC matrix, we apply a log transform $\textstyle(log(SC+1)/max(log(SC+1)))$ (see Fig.~\ref{fig:ResFigure1}A). Then, we use the multi-scale community detection method to determine whether the SC matrix displays hierarchical community structure, and if it does, to characterize that structure both qualitatively and quantitatively. 

In single subjects, we observe that the SC graphs display hierarchical organization where communities branch into smaller sub-communities across a range of topological scales (Fig.~\ref{fig:ResFigure1}B). These characteristics of single-subject SC graphs are recapitulated at the group level. By performing consensus clustering, we can estimate a hierarchical decomposition that is most characteristic of all subjects within the group. We observe a similar hierarchical community structure, indicating a high degree of similarity (and low variance) across subjects (Fig.~\ref{fig:ResFigure1}C). In the Appendix Figs.~\ref{fig:SI_GCSC} and \ref{fig:SI_GCSC_dis}, we show that these group-level communities tend to be composed of spatially proximal, or connected, regions indicating that topological clustering is nontrivially related to spatial location. An interesting exception to this general trend is the existence of a few spatially distributed communities located in the fronto-striatal circuitry that bridge frontal cortex and the striatum. 

The hierarchical community organization in the SC graphs can be described by characteristic curves of community number and size as a function of resolution. Specifically, as the structural resolution parameter value increases, we observe a rapid increase in the number of non-singleton communities {(maximum average number of SC communities = 143.20 $\pm$ 7.3 std)} and an analogous drop in the average community size (Fig.~\ref{fig:ResFigure1}D). As the value of $\gamma$ increases farther, the number of communities branching into singletons increases, and thus the number of non-singleton communities decreases. Together, these trends indicate that communities begin to branch at low $\gamma$ values, and continue to branch as $\gamma$ increases {(full width at half max = 42.49 $\pm$ 2.9, Fig.~\ref{fig:ResFigure1}D)}. Far from haphazard, this global branching process is highly structured, with a large number of nodes maintaining their allegiance to hierarchical communities over a long range of $\gamma$ before branching (see Fig.~\ref{fig:ResFigure1}B).

\subsection{Hierarchical community organization of functional connections in the human brain}

The functional connectivity (FC) values were calculated based on the average wavelet coherence between regional time series, and unlike the SC matrix values, they exhibit a normal distribution with an average value of $\approx 0.49$ and relatively small standard deviation $(\approx 0.04)$ (Fig.~\ref{fig:SI_Fig_edgedist}. C$\&$D). Intuitively, this narrow range of edge weights means that FC graphs will display hierarchical organization over a smaller range of $\gamma$ values. To better sample the FC hierarchical structure, we therefore assessed community organization over a smaller range of $\gamma$ values, with $\gamma \in [0.95, 1.7]$ in the FC matrices as opposed to the $\gamma in [0.0133, 1]$ used for the SC matrices. Importantly, these ranges were chosen pragmatically so as to map out the entire curve from partitions containing a single community (lowest value in the $\gamma$ range), to partitions containing only singletons (highest value in the $\gamma$ range). 

\begin{figure*}
\centering
\includegraphics[width=1\linewidth]{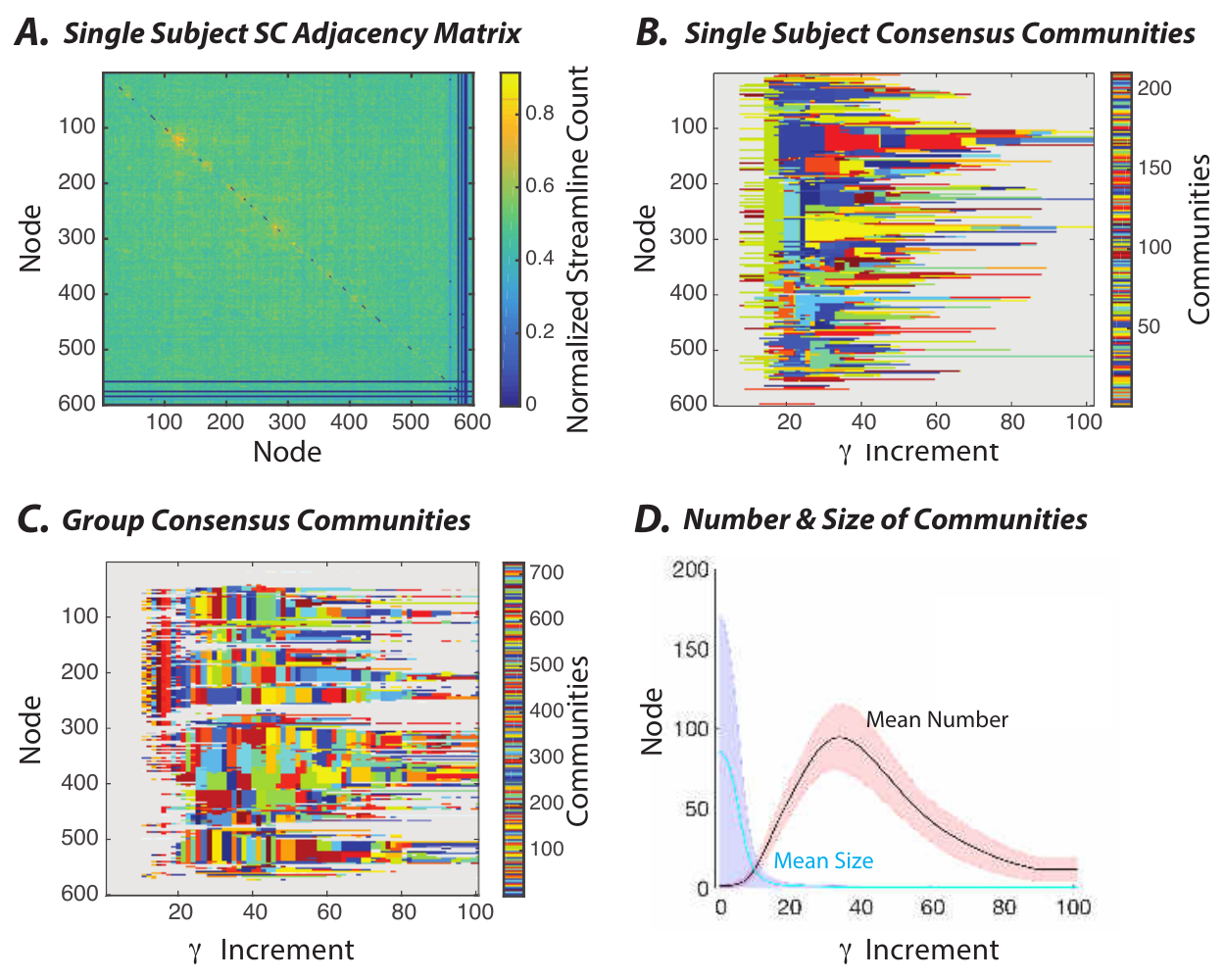}
\caption{\textbf{Application of multi-scale community detection to subject-level and group-level resting state functional brain networks}. \emph{(A)} An example rsfMRI connectivity matrix from one subject, in which each element linking a pair of brain regions represents the pairwise wavelet coherence between regional time series. \emph{(B)} The consensus partition representing the multi-scale community structure for the matrix in panel \emph{(A)}. To enhance the visual detection of the communities, we have represented all singleton communities with the same gray color. \emph{(C)} The consensus partition representing the multi-scale community structure for the group-level functional matrix, which is defined as the average connectivity matrix across participants. Here, again, to enhance clarity, we color the singleton communities in the same gray color. \emph{(D)} The average number as well as the average size (expressed as the percentage of total nodes) of non-singleton communities calculated across layers (i.e., across $\gamma$ increments).  In these analyses, we used a structural resolution parameter $ \gamma \in [ 0.95, 1.7] $, an inter-layer $\gamma$ increment of $= 0.01$, and 75 layers.}
\label{fig:ResFigure5FC}
\end{figure*}

In single subjects, we observe that the FC graphs display hierarchical organization where communities branch into smaller sub-communities across a range of topological scales (Fig.~\ref{fig:ResFigure5FC}B). Although multi-scale communities can be detected reliably in single subjects, the group-level consensus reveals less robust hierarchical organization (Fig.~\ref{fig:ResFigure5FC}C). Indeed, at the group level, communities extend over smaller $\gamma$ ranges before branching (note the speckled nature of the community allegiance matrix). This observation indicates that there is relatively low inter-subject similarity of the hierarchical communities observed in FC graphs.  In fact, a large number of brain regions (especially subcortical regions) fail to display any significant intra-layer community allegiance over most topological scales (see Appendix Figures \ref{fig:SI_GSCFC} and \ref{fig:SI_GSCFC_single}). 

The hierarchical community organization in the FC graphs can be described by characteristic curves of community number and size as a function of resolution. Consistent with the trends observed in the SC graphs, as the structural resolution parameter value increases, we again observe a rapid increase in the number of non-singleton communities (maximum average number of FC communities = 94.44 $\pm$ 20.66 std) and an analogous drop in the average community size (Fig.~\ref{fig:ResFigure5FC}D). As the value of $\gamma$ increases farther, the number of communities branching into singletons increases, and thus the number of non-singleton communities decreases. Again, the global branching process is highly structured, with a large number of nodes maintaining their allegiance to hierarchical communities over a long range of $\gamma$ before branching. Nevertheless, in comparison to the SC graphs, the FC graphs display this hierarchical community organization over a smaller range of $\gamma$ values (full width at half max = 28.86 $\pm$ 7.1, Fig.~\ref{fig:ResFigure1}D). 

\subsection {Heterogeneity in the hierarchical community organization of functional and structural brain networks}

Next we aim to explicitly characterize similarities and differences in the hierarchical community organization of functional and structural brain networks. We begin by focusing on the notion of community longevity or stability across topological scales, and we estimate the average number of stable communities that each node belongs to across layers. Such a computation depends on first choosing a mathematical definition of what constitutes a ``stable'' community. Pragmatically, we choose a parametric definition in which a stable community is defined as a community that exists across more than $x$\% of the $\gamma$ range studied. We refer to the $x$\% as a \emph{stability threshold}. Intuitively, graphs with pronounced multi-scale hierarchical organization display many stable communities per node across a wide range of stability thresholds, while graphs with weak multi-scale hierarchical organization display very few stable communities per node.  In addition to the number of stable communities, it is also of interest to quantify the variance of this number across nodes in the network. Consequently, graphs with a large variance of these values across nodes are characterized by greater topological heterogeneity than graphs with a smaller variance of these values across nodes. 

\begin{figure*}
\centering
\includegraphics[width=0.9\linewidth]{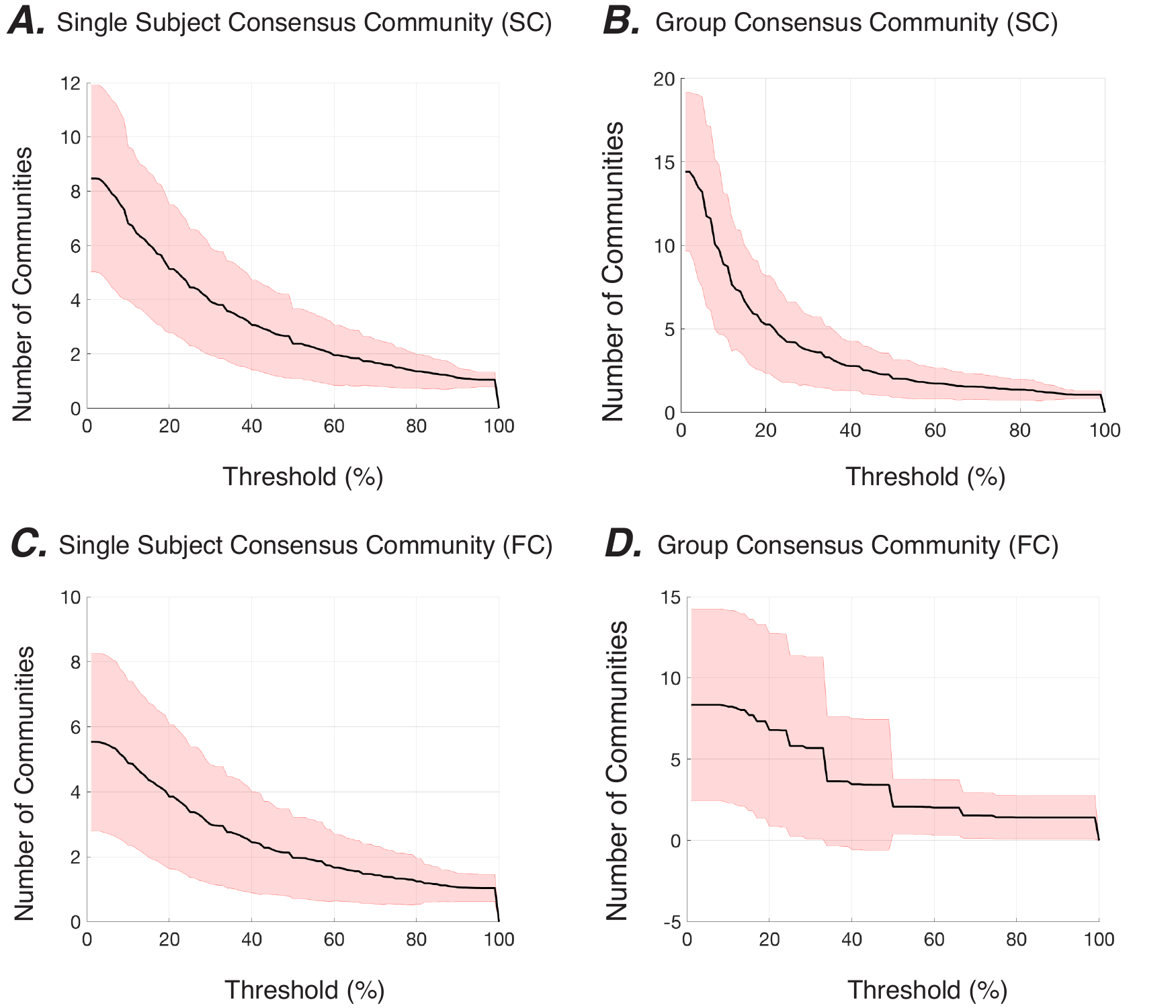}
\caption{\textbf{Local topological scales of hierarchical community organization in structural and functional brain graphs}. The black lines show the average number of communities to which each node is allied across different stability thresholds in the hierarchical community organization of SC \emph{(A-B)} and FC \emph{(C-D)} graphs, at both the subject level \emph{(A,C)} and the group level \emph{(B,D)}. We define the stability threshold as the percent range of $\gamma$ values over which a node is stably allied to a given community, here shown along the $x$-axis. }
\label{fig:ResFigure4}
\end{figure*}

Applying these analyses and statistics to graphs extracted from imaging data, we observe that the brain's structural and functional connectivity graphs are indeed hierarchical, with the vast majority of nodes displaying stable allegiance to communities over more than one topological scale (Fig.~\ref{fig:ResFigure4}). First considering only SC graphs, we observe that at low stability thresholds, each brain region is allied to approximately $8$ communities across topological scales, while at higher stability thresholds, a brain region may only be allied to $1$ community. Across brain regions, we also observe high variance; at low stability thresholds, the number of communities to which a region allies ranges from approximately $5$ to approximately $12$, indicating a high degree of heterogeneity in the SC graphs. Finally, we observe a marked similarity between the community stability curves at the subject level and at the group level (Fig.~\ref{fig:ResFigure4}A$\&$B), providing further evidence of inter-subject similarity of hierarchical community structure in SC graphs. 

Next, considering the FC graphs, we observe that brain regions exist in a smaller number of stable communities across layers (Fig.~\ref{fig:ResFigure4}C$\&$D). The comparison between the FC and SC subject-level results demonstrate that although the average curves appear similar in shape, the average number of stable communities is approximately 1.5 times higher in SC than FC graphs. These observations suggest that FC graphs display a flatter hierarchical community organization, characterized by a smaller number of topological scales. Across brain regions, we again observed relatively high variance; at low stability thresholds, the number of communities to which a region allies ranges from approximately $3$ to approximately $8$, indicating a high degree of heterogeneity in the FC graphs. The community stability curves also display a marked difference at the subject and group levels, again providing evidence of inter-subject variability of hierarchical community structure in FC graphs.

\subsection{Reliable detection of a region's consistent allegiance to communities across topological scales}
 
In the previous sections, we first observed the hierarchical nature of community structure in structural and functional brain graphs, and then we determined the number of topological scales that characterize each node's community allegiance profiles. In this section, we seek to better understand the fine-scale features of the multi-scale network model that allow for reliable estimation of these topological scales in individual brain regions. Importantly, the multi-scale network model is not simply an agglomeration of weighted adjacency matrices. Rather, it explicitly stitches these matrices together with inter-layer connections that link a brain region in one layer to itself in the preceding and following layers. These inter-layer links allow for the quantitative assessment of communities across layers in a statistically principled manner. 

\begin{figure*}
\centering
\includegraphics[width=1\linewidth]{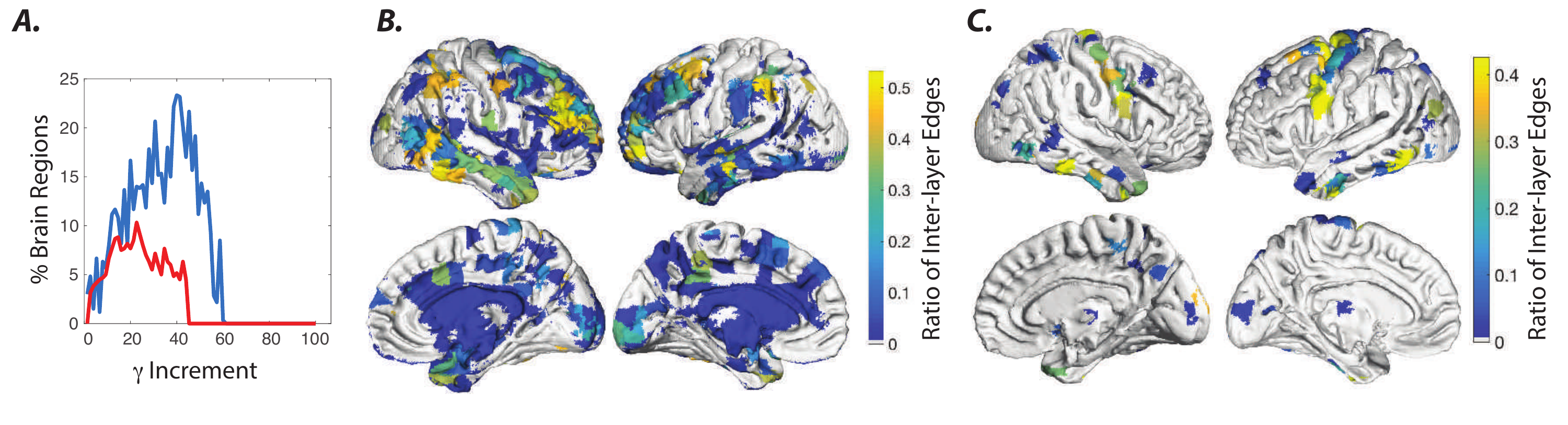}
\caption{\textbf{Reliability of inter-layer multi-scale community allegiance of brain regions in SC and FC graphs}.{ \emph{(A)} The percentage of brain regions from the SC (blue) and FC (red) graphs containing inter-layer edges with significant probability at the group-level across $\gamma$ increments. The probability of identifying the SC inter-layer edges fall below the chance level for the second half of the $\gamma$ range values. These results suggest that the estimated stability and the hierarchical structure of the higher $\gamma$ layers are unreliable. Inter-layer coupling weights were low $( \tau = 0.25)$. The probability of identifying the FC inter-layer edges was above the significance level for a small range of $\gamma$ values. Since the vast majority of the group-level consensus inter-layer edges are not significant, the identified communities frequently switch across layers (as seen in (Fig.~\ref{fig:ResFigure5FC}C))}.  \emph{(B-C)} The regions from the SC (panel B) and FC (panel C) graphs containing inter-layer edges with significant probability at the group-level are overlayed on the brain and color-coded to represent the ratio of the significant inter-layer edges (i.e., by dividing by the total number of inter-layer edges). }
\label{fig:Inter_Layer_fig}
\end{figure*}

Explicit inter-layer links within the multi-scale model motivate an effective description of inter-layer consistency in a node's allegiance to communities. In particular, over the large number of optimizations of the modularity quality function that must be performed to adequately sample the underlying landscape, it is of interest to quantify how frequently a node remains in the same community across two adjacent layers (here representative of topological scales). We define a reliable inter-layer association as occurring when a node remains in the same community across two adjacent layers for a greater number of optimizations than expected by chance (see Methods). We observe that a large number of reliable inter-layer associations can be identified in structural brain graphs at both the subject and group levels. This is true particularly for lower values of $\gamma$, indicating the presence of a $\gamma$ range over which hierarchical community assignments can be reliably detected (Fig.~\ref{fig:Inter_Layer_fig}A). While a number of cortical and most of the subcortical structures including dienchephalone and limbic system display reliable inter-layer association across a small range of $\gamma$ increments at the group-level, several bilateral clusters in the frontal, parietal, and temporal cortex display a notably higher range (Fig.~\ref{fig:Inter_Layer_fig}B). In FC graphs, we observe reliable inter-layer associations over a much smaller range of $\gamma$ at the subject level (Fig.~\ref{fig:Inter_Layer_fig} A). At the group level, we not only observe that reliable inter-layer associations occur over a small range of $\gamma$, but also that there are very few reliable associations at all (Fig.~\ref{fig:Inter_Layer_fig}C). Together, these findings underscore both the flatter hierarchical nature of FC graphs and the greater inter-subject variability in comparison to SC graphs. Yet, the identified regions from the SC(FC) graphs with reliable inter-layer associations is a reflection of the fact that networks of structures with similar and/or strong structural (functional) connectivity are present across subjects.


\subsection{Homogeneity \emph{versus} heterogeneity of topological scales in structural and functional brain graphs}

The previous results indicate that we can reliably detect hierarchical community structure in structural and functional brain graphs, and that the two types of graphs display differing degrees of topological heterogeneity. To better understand this heterogeneity, particularly across nodes in the network (or regions in the brain), we examine the stability of communities more closely. Specifically, for node \textit{i}, we measure the stability of its allegiance to community \emph{`X'} by calculating the fraction of layers in which node \textit{i} belongs to community \emph{`X'} (see Methods). This formulation allows us to define a stability matrix by replacing the community label with the stability of the node's allegiance to the community (Appendix Fig.~\ref{fig:ResFigure3}A-B). This matrix quantifies the stability of a node at a given structural resolution parameter value. In this matrix, highly diverse patterns of community allegiance stability are indicative of topological heterogeneity in the graph. By contrast, less diverse patterns of community allegiance stability are indicative of topological homogeneity in the graph.

To quantity homogeneity \emph{versus} heterogeneity, we decomposed the stability matrix using a principle component analysis, such that each component indicated a coherent pattern of communities across scales (Appendix Fig.~\ref{fig:ResFigure3}C-D). Intuitively, graphs with greater topological heterogeneity require a larger number of principle components to explain a given amount of variance in the community stability matrix compared to more homogeneous graphs. In both SC and FC graphs, we observed that a handful of components explained most of the variance in the community stability matrix (Appendix Fig.~\ref{fig:ResFigure3}E-F). In SC graphs, eight principle components explain more than $95\%$ of the variance in the stability matrix. The first component is marked by the stability profile of the singleton nodes, and the second component highlights the lower half of the $\gamma$ range where most of the larger communities reside. In FC graphs, only five principle components explain more than $95\%$ of the variance in the stability matrix. The group-level analysis shows that on average significantly ($t$-test, $p < 0.001$) smaller number of principle components ($4.86 \pm 0.78$) explain subjects FC stability matrices compared to that of the SC graphs ($9.49 \pm 1.26$). These results provide converging evidence that SC graphs display a greater topological heterogeneity while FC graphs display a greater topological homogeneity in hierarchical community structure.

\subsection{Comparison between the hierarchical community organization of the brain's structural and functional connectivity}
 
In previous sections, we demonstrate that the hierarchical organization of the SC and FC graphs differ both in terms of the presence and stability of communities across topological scales. Yet it is also important to note that the identified communities across modalities do share some similarities, perhaps supporting the notion that structure provides the scaffold for emergent functional dynamics. To better understand the similarities between the hierarchical community organization of SC and FC graphs, we begin by summarizing the community structure at each $\gamma$ value in each modality as an $N \times N$ allegiance matrix, where the $ij^{th}$ element indicates the fraction of times that node $i$ and node $j$ are placed in the same community over all optimizations of the multilayer modularity quality function. 

\begin{figure*}
\centering
\includegraphics[width=1\linewidth]{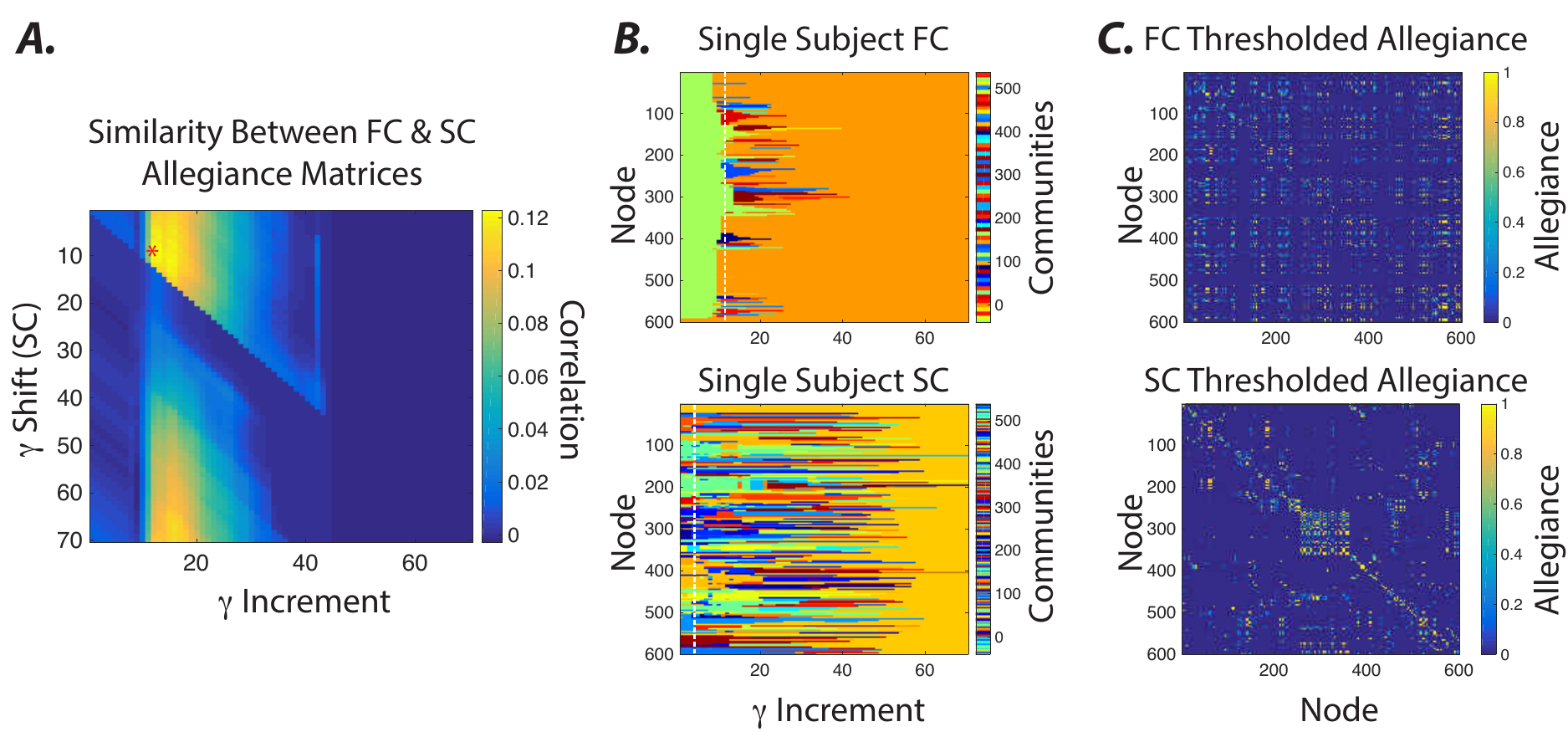}
\caption{\textbf{Similarity between the hierarchical community structure of structural and functional brain graphs}.  \emph{(A)} The similarity between the allegiance matrices of the hierarchical community structures of the FC and SC graphs. Specifically, we calculate the Pearson correlation coefficient between the allegiance matrix of SC at a given $\gamma$ and the allegiance matrix of FC at a given $\gamma$, for all possible $\gamma$ pairs, allowing us to identify the alignment that results in the highest similarity between their allegiance matrices (marked by $\color{red} * \color{black}$ in the plot). The dashed white lines in \emph{(B)} highlight the layers where the SC and FC realignment yield highest similarity values. \emph{(C)} The thresholded allegiance matrices of layers highlighted by dashed lines in \emph{(B)} of the FC and SC hierarchical communities. FC and SC allegiance matrices are order identically based on their original node labels. }
\label{fig:ResFigure5}
\end{figure*}

Next, we calculate the Pearson correlation coefficient between the allegiance matrix of SC at a given $\gamma$ and the allegiance matrix of FC at a given $\gamma$, for all possible $\gamma$ pairs (Fig.~\ref{fig:ResFigure5}A). This approach enables us to capture the degree to which densely connected communities of brain regions in FC similarly echo their underlying SC, thereby providing insight into the structural drivers of global dynamics. We observe that SC and FC communities show high similarity at medium topological scales (Fig.~\ref{fig:ResFigure5}B), suggesting that it is not simply the case that densely structurally connected regions are also functionally connected. Instead, these results suggest that medium-sized bundles that link the densely connected (and commonly local) brain regions allow global functional synchronization between relatively large ensembles (Fig.~\ref{fig:ResFigure5}C).

\subsection{Explicit multimodal investigation using a multiplex, multi-scale graph}

While the comparisons thus far between structural and functional brain graphs have been illuminating, it is natural to ask whether there is a more principled and model-based approach to comparing the two modalities within the multilayer framework. Indeed, the multilayer framework does allow additional graphs to be interconnected along distinct dimensions. Thus, it is possible to construct a graph where one dimension hard-codes topological scale (as done throughout the earlier sections of this paper), and a second dimension that hard-codes imaging modality (e.g., structural connectivity and functional connectivity). 

Here we construct exactly this multiplex graph to more formally study the multi-scale nature of both the structural and functional connectivity matrices within the same model (Appendix Fig.~\ref{fig:ResFigure6}A). In addition to inter-scale links $\tau$, this model also contained inter-modality links $\kappa$ that link a node in one scale and one modality to itself in the same scale in a different modality. We optimize the modularity quality function in this multiplex case to identify the hierarchical community structure of the SC and FC graphs. Importantly, as $\kappa$ is tuned down, community structure is allowed to be independent in the SC and FC graphs (Appendix Fig.~\ref{fig:ResFigure6}B). In contrast, when $\kappa$ is tuned up, community structure is forced to be consistent across the two types of graphs (Appendix Fig.~\ref{fig:ResFigure6}F). In other words, by employing higher values of $\kappa$, we are able to extract community structure that is most representative of the graphs in both imaging modalities. 

Interestingly, we observe that this cross-modality community structure appears more similar to the community structure of the SC graphs when they were studied independently, than to the community structure of the FC graphs when they were studied independently. This phenotype can occur in community detection when the community structure in one graph is stronger than the community structure in the other graph. To investigate and more thoroughly quantify this observation, we study the similarity between the allegiance matrices of the multiplex SC-FC graph, the FC graph alone, and the SC graph alone, as a function of the topological scale ($\gamma$ value) at which they were constructed (Appendix Fig.~\ref{fig:ResFigure7}). We observe that the hierarchical structure of the multiplex SC-FC graph is similar to that of the FC graph alone only in a narrow range of topological scales, consistently across $\kappa$ values (Appendix  Fig.~\ref{fig:ResFigure7}A-B). In contrast, we observe that the hierarchical structure of the multiplex SC-FC graph is similar to that of the SC graph alone across a wide range of topological scales, and consistently across $\kappa$ values (Appendix Fig.~\ref{fig:ResFigure7}C-D). These results suggest that the joint optimization is more heavily influenced by the hierarchical community structure in the SC graph than it is by that of the FC graph. 

Importantly, these results are reported over a single subject, and thus it is critical to ask to what degree these insights hold over the entire participant cohort. To address this question, we perform the same set of analyses, but instead of using the single-subject allegiance matrices, we use the group-level allegiance matrices. In general we observe consistent results at the group scale. Specifically, the hierarchical community structure of the muliplex SC-FC graphs at high $\kappa$ values are consistently reminiscent of the SC structure (highest observed correlation approximately $r=0.85$), along a range of topological scales (Fig.~\ref{fig:ResFigure8}A). And they are reminiscent of the FC structure to a weaker degree (highest observed correlation approximately $r=0.41$), along a much narrower range of topological scales (Fig.~\ref{fig:ResFigure8}B). Notably, the similarity between either SC or FC and the multiplex SC-FC graphs is higher than between SC and FC alone (highest observed correlation approximately $r=0.25$). The FC and SC communities at coarse topological scales show the highest similarity (Fig.~\ref{fig:ResFigure8} C-F), however they diverge at higher topological scales as the SC communities branch into smaller local communities (Appendix Fig.~\ref{fig:SI_GCSC}), whereas the FC communities at higher topological scales are more spatially distributed. These results confirm at the group level that the joint optimization is more heavily influenced by the hierarchical community structure in the SC graph than it is by that of the FC graph. In addition, regions that display comparable community allegiance between FC and SC graphs such as the subcortical nodes and some clusters within the medial frontal and medial occipital cortices are also identifiable in the multiplex SC-FC graph's multi-scale communities as they maintain their community allegiance across $\gamma$ increments. Overall the observed similarity between the multiplex SC-FC graph's communities and the FC and SC communities provides converging evidence that both modalities share major organizational features. Nevertheless the multiplex SC-FC graph's communities are comparable to the original FC and SC communities at different hierarchical scales, which highlights the differences in the hierarchical community organization of the SC and FC graphs.

\begin{figure*}
\centering
\includegraphics[width=1\linewidth]{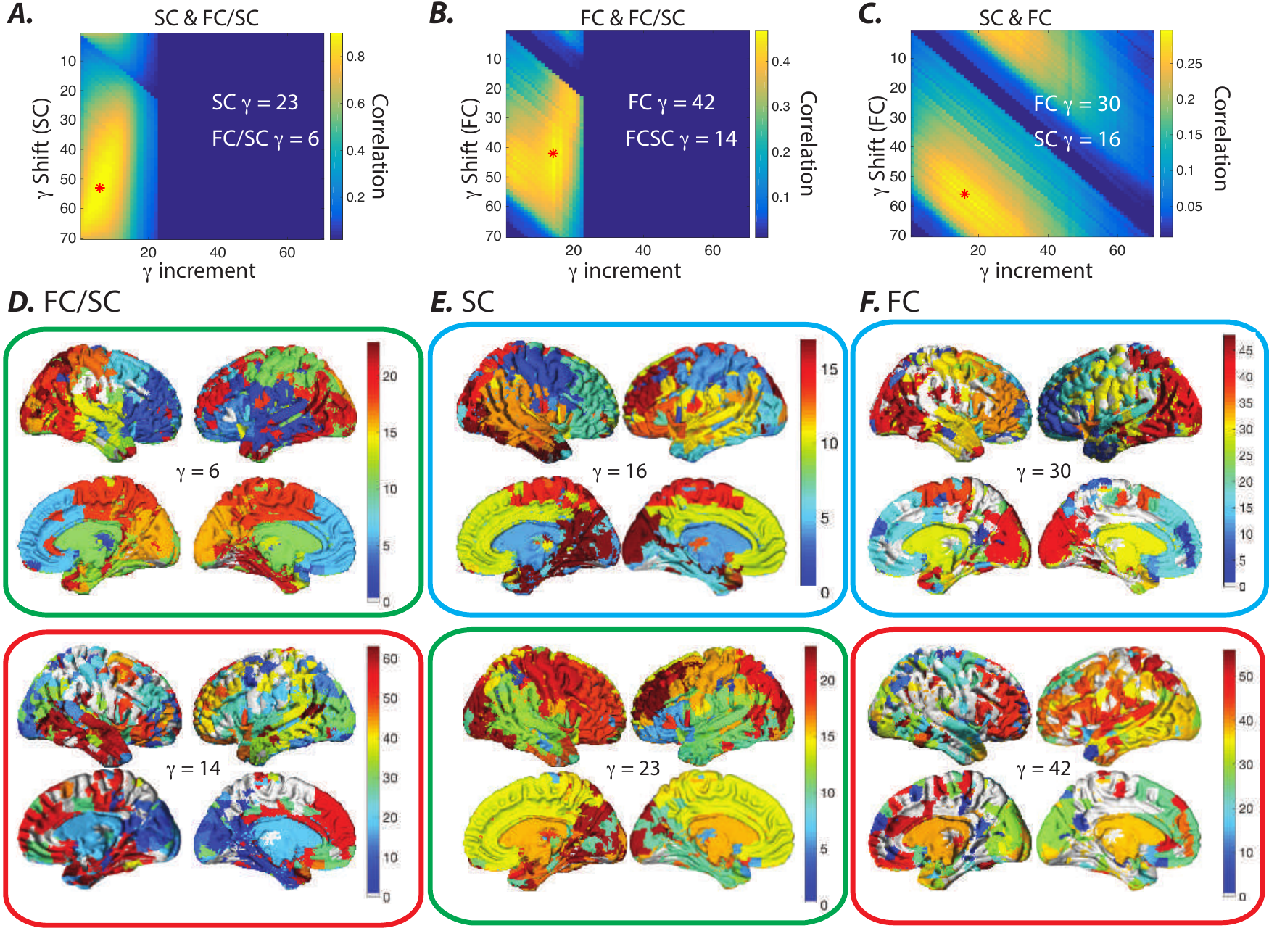}
\caption{\textbf{Similarity between the allegiance matrices of the multiplex SC-FC graph, the FC graph alone, and the SC graph alone, as a function of the topological scale for the group}. Using a $(\kappa = 1)$, we calculated the Pearson correlation coefficient values for all layers and all $\gamma$ shifts between the allegiance matrices of \emph{(A)} the multiplex SC-FC graph and the SC graph, \emph{(B)} the multiplex SC-FC graph and the FC graph, and \emph{(C)} the SC graph and the FC graph outside of the multiplex formulation. Layers with maximum correlation values are marked by $\color{red} * \color{black}$ \emph{(D-F)}  Brain overlays show the communities identified at layers with maximal correlation values (panels \emph{(A-C)}). The color bars in each panel (right hand side) represents the color-coded communities. The multiplex SC-FC communities at low $\gamma$ layers are very similar to SC communities at slightly higher $\gamma$ layers (green boxes). Although FC communities overall show smaller similarity to the muliplex SC-FC communities, the multiplex SC-FC communities at higher $\gamma$ layers show relatively higher similarity to FC communities at high $\gamma$ layers (red boxes). The smallest similarity is observed between FC and SC graphs, peaking around a course scale with small $\gamma$ values for both modalities (cyan boxes).}
\label{fig:ResFigure8}
\end{figure*}

\section{Discussion}

The human brain is a complex system that can be fruitfully represented as a graph or network in which brain regions correspond to network nodes and structural or functional connections between regions correspond to network edges \citep{bullmore2009complex}. Recent observations have pointed to the fact that both structural and functional brain networks may have community structure \citep{sporns2016modular}: the presence of densely interconnected groups of regions that might support specific cognitive functions \citep{power2011,yeo2011,meunier2009}. Moreover, evidence suggests that these communities exist over multiple topological scales \citep{bassett2010efficient}, with larger communities potentially being composed of smaller communities \citep{betzel2016multi}. Yet a comprehensive characterization of this putative hierarchical community structure in structural and functional brain graphs has remained difficult largely due to inadequacies in existing analytical paradigms and computational tools. Here we address these limitations by exercising a multi-scale community detection algorithm \citep{mucha2010}, and applying it to both structural brain networks estimated from diffusion imaging and functional brain networks estimated from resting state fMRI. Using novel statistics including community stability and inter-scale reliability, we show that structural brain graphs display a wider range of topological scales than functional graphs. We also illustrate the utility of this method in examining multimodal graphs that combine both structural and functional connectivity information. Our work illustrates the practical utility of multi-scale community detection in revealing hierarchical community structure in brain graphs, and opens the door for future investigations of this structure in both health and disease.

\subsection{Detecting multi-scale community structure}

Characterization of multivariate dependencies across spatio-temporal scales is critical for a fundamental understanding of observable dynamics across systems as diverse as the climate system \citep{steinhaeuser2012multivariate} and the human brain \citep{betzel2016multi}. The multi-scale community detection algorithm that we exercise here reveals the hierarchical community organization of a graph by assuming dependence between neighboring topological scales \citep{mucha2010}. A marked advantage of this approach compared to conventional single-scale algorithms is that it provides a statistically principled answer to the question: ``Is a community at one scale the same as or different from a community at another scale.'' Perhaps even more importantly, the approach provides an estimate of the stability of local topological structure, and therefore a pragmatic means of identifying model parameter values that maximize the consistency of locally stable communities across several topological scales. These local estimates of community stability (unlike the global measures of community stability that have been previously defined \citep{delvenne2010stability,pons2005computing,arenas2008analysis,ronhovde2009multiresolution,karrer2008robustness}) are robust to topological heterogeneities \citep{danon2006effect} in the form of communities of different sizes with different average edge weights.  When applied to human brain networks, we find that the local community stability estimates allow characterization of communities that are stable across a range of topological scales. Taken together, our study offers not only a methodological approach to studying hierarchical community structure in graphs, but also a set of statistical methods to characterize the observed structure and to compare it across different classes of graphs, either treated separately or combined into a multiplex model.

\subsection{Multi-scale community structure in the human brain's white matter architecture}

Structural brain graphs estimated from diffusion imaging data tend to be sparse, and the edge weight distributions tend to be heavy-tailed \citep{lohse2014resolving,bassett2011conserved,hagmann2008mapping}. These characteristics can occur when a complex topology is embedded into a 3-dimensional space \citep{bassett2010efficient}, in such a way as to enhance the efficiency of information transmission \citep{bullmore2009generic} while decreasing the cost of the wiring \citep{bullmore2012economy}. Interestingly, prior work has also offered initial evidence that the complexity of structural connectivity is in part due to the fact that it is organized in a hierarchically modular fashion \citep{lohse2014resolving}, which is thought to support its information processing capabilities \citep{simon1991architecture}. Here we use a principled mathematical modeling approach to more exactly identify hierarchical community structure in structural brain graphs. Our results demonstrate that structural connectivity is characterized by heterogeneous multi-scale communities, and by nodes that form stable hierarchical communities across a range of topological scales. Multi-scale communities appear to be largely consistent across different subjects and also tend to be spatially localized. The observation of both regional heterogeneity and diverse topological scales indicates that the application of single-scale community detection techniques is likely to produce an overly-simplified picture of the brain's organization. 

While the majority of multi-scale communities were spatially localized, communities in basal ganglia-thalamo-cortical circuitry were not. Over several topological scales, subcortical structures including basal ganglia (mainly putamen, palladum, and caudate) and anterior thalamus as well as several frontal neocortical areas were identified within the same community. While frontal and subcortical structures are spatially distributed, it is commonly known that much of the cortex (including both allocortex and isocortex) projects to the striatum \citep{swanson2000cerebral}, although not all projections are entirely reciprocal. These consistent projections can manifest as structural motifs that are accessible to community detection algorithms. Other complementary algorithms based on tools from algebraic topology, including the notions of persistent homology \citep{giusti2016twos}, have demonstrated that basal ganglia-thalamo-cortical connections are among the few most common \emph{cycles} identifiable in structural brain graphs \citep{sizemore2016closures}. These cycles and motifs are known to play key roles in rhythmic gain control, and in the gating and integration of information across the brain \citep{womelsdorf2014dynamic, rajan2016recurrent}. For example, basal ganglia influence cortical states and behavior via dopaminergic inputs to thalamus, thereby enabling the integration of information characteristic of reinforcement learning \citep{schultz1997neural,schultz1998predictive}. Indeed, basal ganglia input is modulatory and also serves to gate higher-order relay signals that are propagated through cortico-thalamo-cortical loops \citep{sherman2006exploring}. Together, the unique role that the basal ganglia-thalamo-cortical pathways serve in driving brain states is made possible through the unique structural fingerprint of subcortical regions.

\subsection{Multi-scale community structure in the human brain's resting state functional connectivity}

The pattern of phase-locking between regional BOLD time-series over a period of several minutes demonstrates that many regions display high functional connectivity with one another \citep{yaesoubi2015dynamic}. While the resulting graph is relatively dense and homogeneous, it nevertheless displays some amount of hierarchical community structure. In single subjects, the hierarchical consensus analysis revealed a small range where multi-scale community structure is reliably identified. Interestingly, the group-level consensus analysis showed that the majority of brain regions failed to produce reliable inter-layer links across individuals, indicating the high degree of inter-subject variance in hierarchical community organization. These results are not entirely unexpected in light of the mounting evidence for both inter-session and inter-subject variability in resting state functional connectivity \citep{gonzalez2014spatial}, particularly that located in heteromodal association areas \citep{mueller2013individual,finn2015functional}. Speculatively, it is possible that some of this inter-subject variability is due to the fact that these heteromodal association areas are more susceptible to and likely influenced by environmental factors, a fact highlighted by research on the postnatal period where they display protracted development during a time period of high plasticity \citep{mueller2013individual,brun2009mapping, zilles2013individual}. Thus, the observed inter-subject variability in the multi-scale community structure of functional brain graphs could provide important fodder for a fundamental understanding of the principles of brain wiring, evolution, and ontogenetic development \citep{zilles2013individual}.

\subsection{A comparison of hierarchical community structure in functional and structural graphs}

It has long been observed that resting state functional connectivity shows statistically similar patterns to those observed in underlying structural connectivity \citep{hagmann2008mapping,vincent2007intrinsic,sporns2013structure}. Yet, we have little understanding of how exactly the anatomical connections gives rise to observed functional interactions \citep{hermundstad2013structural,hermundstad2014structurally}. Evidence suggests that the relationship between structure and function is likely quite indirect, with the broader network adjacent to direct structural paths being critical to healthy dynamic couplings between brain regions \citep{goni2014resting,becker2015accurately}.  Our work supports this notion by demonstrating that the highest similarity in hierarchical community structure between the two modalities is found between the fine topological scales of the functional graph and the relatively coarse topological scales of the structural graph, which takes into account broader anatomical network organization. The differential scales of function and structure that map onto one another can in part be explaind by the observation that structural graphs on average display $\approx 1.5$ times more topological scales than functional graphs. Together these results suggest that (i) functional connectivity dynamics are not strictly bound to or constrained within direct anatomical projections, but instead extend to spatially distributed circuits, and (ii) structural connections as estimated by white matter tractography display hierarchical community structure that can support long range functional coupling \citep{werner2009fractals,valverde2015structural}. 

It is important to note that although we found similarities between the hierarchical community structure of functional and structural graphs at different topological scales, the overall magnitude of the similarity was relatively small. One fundamental property of brain connectivity that might explain this relative independence of structural and functional graphs is temporal dynamics \citep{mattar2016flexible,hutchison2013dynamic,bassett2011,preti2016dynamic}. Indeed, while we have here studied a static functional graph that represents patterns of co-activation over several minutes, in reality the brain displays time-varying patterns of functional connectivity \citep{calhoun2014chronnectome} that can track changes in cognitive processes \citep{braun2015dynamic,mattar2015functional,chai2016functional} and behavior \citep{vatansever2015default,gerraty2016dynamic,bassett2015learning}. The nature of these dynamics suggest that the answer to the question ``how does structure constrain function?'' depends nontrivially on the time scale of the function (or functional connectivity pattern) in question. Indeed, recent work has demonstrated that synchronization patterns in hierarchical modular structures such as the brain may appear at different topological scales depending on the time scale of their interaction \citep{arenas2006synchronization,villegas2014frustrated,betzel2013multi,betzel2016multi,honey2007network,PhysRevLett.97.238103}. Therefore, comparing the topology of the multi-scale functional and structural graphs can prove useful for understanding the properties of anatomical projections that are critical for the emergence of multi-scale functional patterns.

\subsection{Explicit multimodal models of the brain's hierarchical community structure}

While the method that we propose and exercise is applicable to brain graphs constructed from a single imaging modality, it is also flexible and generalizable to questions that require fusion of brain graphs constructed from two or more modalities. We illustrate the method's utility in this class of problems by exploring the joint optimization of modularity across a multiplex network composed of both functional and structural layers \citep{kivela2014multilayer}. This construction enables us to highlight the characteristics of community structure that are echoed across the two imaging modalities: diffusion imaging and resting state functional MRI. Our results reveal that at high inter-modal coupling strengths, the community organization of the multi-modal graph merges across modalities to form a hybrid structure. Similarity analysis revealed that the multi-modal graph shows highest similarity to functional graphs at relatively fine topological scales and to structural graphs at relatively broad topological scales. The narrow range of topological scales at which this hybrid structure was identified highlights the fact that community structure in functional dynamics extends beyond direct white matter projections. This work complements prior efforts to bridge functional and structural connectivity patterns using tools and techniques that span the purely qualitative and the exquisitely quantitative: these approaches include direct superposition, fiber tracking from functional parcels, and regression analysis relating functional, anatomical, and behavioral data \citep{rykhlevskaia2008combining}. One recent study identified communities separately for each modality, and then subsequently maximized a cross-modularity function to identify a community partition shared by structure and function at medium-to-fine topological scales \citep{diez2015novel}. Our work extends these findings by assessing hierarchical community organization in a multiplex network composed of both functional and structural layers. Future work could seek a better understanding of how joint optimization of graphs with varying topologies across modalities can lead to the robust detection of shared features.

\subsection{Methodological considerations}

Several methodological considerations are pertinent to this work. First, although the multi-scale community detection algorithm identifies the entire hierarchical community structure simultaneously, the resolution at which we study the topology of the graph is limited by the number of layers in the multi-scale graphs. This means that for very large graphs, high resolution multi-scale community detection can be computationally intensive. This property can be especially limiting in multi-modal graphs where two graphs show widely different hierarchical organizations, thus making it time-intensive to accurately sample the modularity landscape. Second, one of the greatest advantages of using conventional community detection algorithms is that they provide a single partition of nodes into communities. Using this partition we can extract additional information regarding the role that individual nodes play within the graph by calculating summary statistics including the participation coefficient and within-module degree $z$-score \citep{guimera2005functional}. However, there are currently no equivalent summary statistics for multi-scale community organization, and therefore the utility of this approach could be significantly enhanced by the parallel development of such statistics. Third, the modularity quality functions that we studied in this work incorporate a uniform null model \citep{bassett2013robust, wymbs2012differential, bassett2015extraction, papadopoulos2016evolution} instead of the more traditional Newman-Girvan null model \citep{newman2010networks}. We chose this null to preserve the one-to-one relationship between the stability of the local communities and the absolute edge values. We observed that the Newman-Girvan null model changes this relationship, thereby altering the relative estimates of local stability. Future work should explore different null models for multi-scale and multi-modal communities \citep{paul2016null}, with the goal of determining their relative utility in extracting hierarchical community structure from brain graphs. Finally, it is worth noting that due to computational limitations, we only examined low values of the inter-layer (scale) dependence. Future work could extend our observations by assessing hierarchical community structure apparent at different values of the inter-layer weight, or when linking layers either ordinally or categorically \citep{mucha2010}.

\section{Conclusion}

In this work, we examined a multi-scale community detection algorithm and its advantages for uncovering the hierarchical organization of synthetic and real world graphs. By assuming dependence between the adjacent topological scales, the multi-scale algorithm links the communities persisting over several scales, thereby effectively uncovering hierarchical community organization in graphs. We demonstrated the statistical robustness of this hierarchical organization by defining notions of community stability and inter-scale reliability. After exercising the method on synthetic graphs, we next examined and compared the hierarchical community organization of structural brain graphs and functional brain graphs estimated from diffusion imaging and resting state functional MRI, respectively. Compared to the functional graphs, the structural graphs displayed a higher degree of topological heterogeneity with a more pronounced hierarchical organization as evidenced by a higher average number of stable communities across topological scales. With the exception of basal ganglia-thalamo-cortical circuitry, the structural communities across topological scales tended to be spatially localized, where nodes within a community were located in close physical proximity to one another. Interestingly, functional communities displayed weak similarity to structural communities at coarse topological scales, and this dissimilarity became more pronounced at finer topological scales as spatially distributed functional communities emerged. These statistical differences between the spatially distributed functional communities and spatially localized structural communities were also apparent in an explicit multi-modal extention of our method, which performs a joint optimization of modularity across a multiplex network composed of both functional and structural layers. Taken together, these results illustrate the practical utility of multi-scale community detection in revealing hierarchical community structure in single-modality and multi-modality brain graphs.

\section{Acknowledgments}
This work was supported by the Army Research Laboratory through contract number W911NF-10-2-0022. DSB would also like to acknowledge support from the John D. and Catherine T. MacArthur Foundation, the Alfred P. Sloan Foundation, the Army Research Office through contract number W911NF-14-1-0679, the National Institute of Health (2-R01-DC-009209-11, 1R01HD086888-01, R01-MH107235, R01-MH107703, R01MH109520, 1R01NS099348 and R21-M MH-106799), the Office of Naval Research, and the National Science Foundation (BCS-1441502, CAREER PHY-1554488, BCS-1631550, and CNS-1626008).The content is solely the responsibility of the authors and does not necessarily represent the official views of any of the funding agencies.

\label{}

\appendix

\section{Spectral community retection of rynamic (multi-slice) networks}

The single layer modularity quality function has been generalized to multi-slice networks to identify communities in multiplex or time-dependent networks. Formally, the multi-slice modularity quality function can be defined as   

\begin{equation}
Q = \frac{1}{2\mu }\sum_{ijlr} \left \{ \left ( A_{ijl}-\gamma_{l} P_{ijl} \right )\delta_{lr} +\delta_{ij}\omega_{jlr}  \right \}\delta \left ( g_{il},g_{jr} \right )\\,
\end{equation}

\noindent where the adjacency matrix of layer $l$ has components $A_{ijl}$, the element $P_{ijl}$ gives the components of the corresponding layer-$l$ matrix for the null model, $\gamma_{l}$ is the structural resolution parameter of layer $l$, $g_{il}$ gives the community assignment of node $i$ in layer $l$, $g_{jr}$ gives the community assignment of node $j$ in layer $r$,  $\omega_{jlr}$ gives the connection strength (i.e., an \emph{inter-layer coupling} parameter) from node $j$ in layer $r$ to node $j$ in layer $l$, the total edge weight in the network is $\mu = \frac{1}{2}\sum_{jr} K_{jr}$, the strength (i.e., weighted degree) of node $j$ in layer $l$ is $K_{jl}= k_{jl} + C_{jl}$, the intra-layer strength of node $j$ in layer $l$ is $k_{jl} = \sum_{i} A_{ijl}$, and the inter-layer strength of node $j$ in layer $l$ is $k_{jl} = \sum_{r} \omega_{jlr}$.

Here we extend the multi-scale framework to multi-slice networks. Formally the multi-slice mulit-scale modularity quality function we study can be defined as   

\begin{equation}
Q = \frac{1}{2\mu }\sum_{ijlrxy} \left \{ \left ( A_{ijl}-\gamma_{l} P_{ijl} \right )\delta_{(lx,ry)} +\delta_{(ix,jy)}\omega_{jlrx} +\delta_{(il,jr)}\tau_{jxy}  \right \}\delta \left ( g_{ilx},g_{jry} \right)\\,
\end{equation}

\noindent where the adjacency matrix of layer $l$ has components $A_{ijl}$, the element $P_{ijl}$ gives the components of the corresponding layer-$l$ matrix for the null model, $\gamma_{l}$ is the structural resolution parameter of layer $l$, $g_{il}$ gives the community assignment of node $i$ in layer $l$, $g_{jr}$ gives the community assignment of node $j$ in layer $r$,  $\omega_{jlrx}$ gives the connection strength  from node $j$ in layer $r$ to node $j$ in layer $l$ at scale layer $x$,  $\tau_{jxy}$  gives the \emph{topological scale coupling parameter} which indicates the strength of the links between neighboring topological scales (as represented by layers), from node $j$ in scale layer $x$ to node $j$ in scale layer $y$. the total edge weight in the network is $\mu = \frac{1}{2}\sum_{jrx} K_{jrx}$, the strength (i.e., weighted degree) of node $j$ in layer $l$ and scale layer $y$ is $K_{jly}= k_{jly} + C_{jly} + T_{jly}$, the intra-layer strength of node $j$ in layer $l$ and scale layer $y$ is $k_{jly} = \sum_{i} A_{ijl}$, and the inter-layer strength of node $j$ in layer $l$ and scale layer $y$ is $C_{jly} = \sum_{r} \omega_{jlry}$, and the inter-scale strength of node $j$ in layer $l$ and scale layer $y$ is $T_{jly} = \sum_{x} \tau_{jlxy}$. 

Finally we provide a synthetic example in Fig.~\ref{fig:Figure4} to show how the multi-scale community detection algorithm links communities across temporal scales and to uncover the relationships between them.  

\begin{figure*}
\centering
\setcounter{figure}{0}    
\includegraphics[width=1\linewidth]{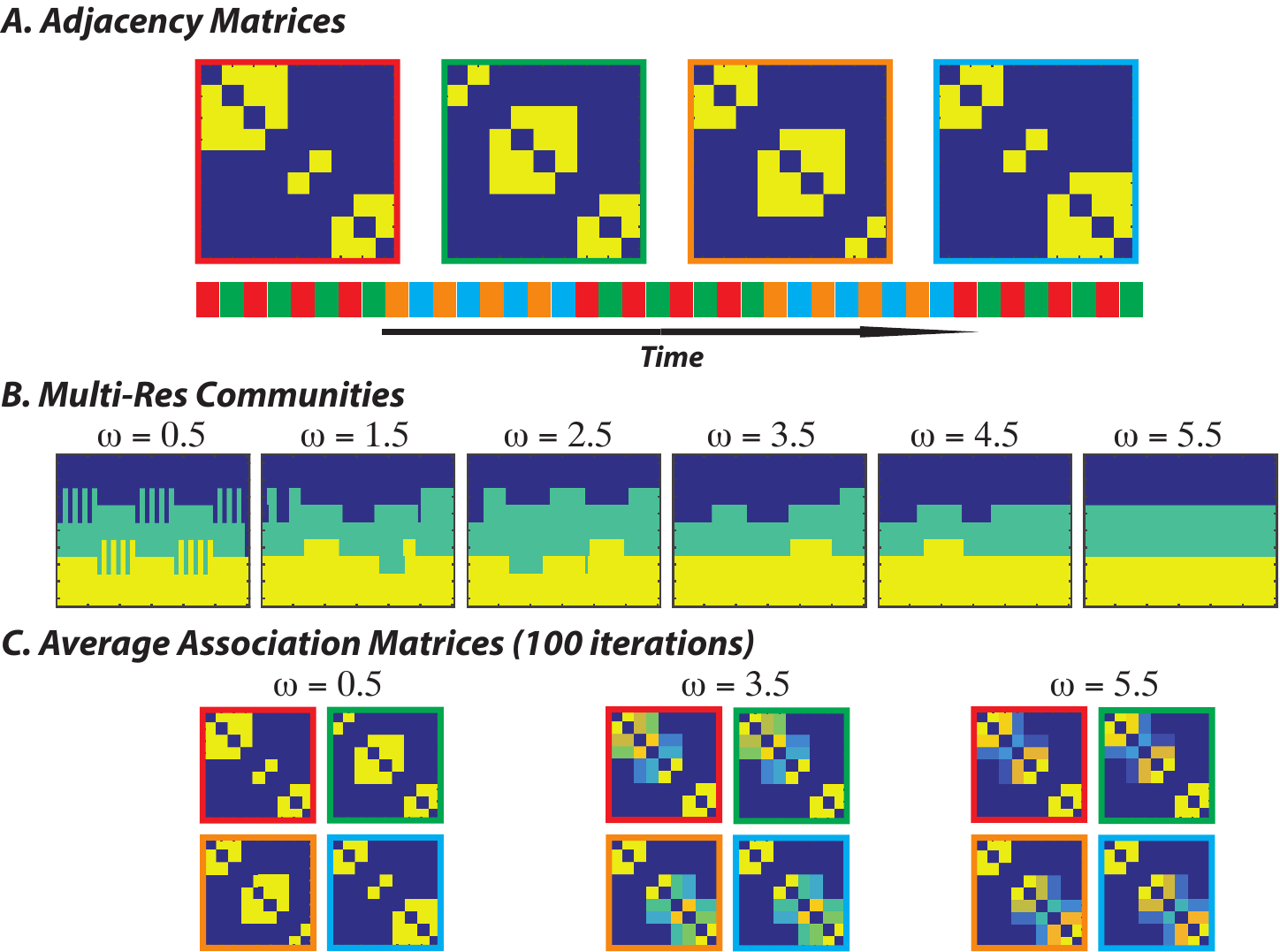}
\caption{\textbf{Multi-scale community organization of a synthetic dynamic graph}. \emph{(A)} We created a dynamic graph by periodically changing the community structure of the graph. In this example, the network starts by switching between two community structures, indicated by the colors red and green. However, at slower time scales the switching dynamics between the red and green community structures periodically changes to an alternative switching dynamics between two different community structures, indicated by blue and orange. In this way, we create dynamics in the graph's community structure at two different time scales. \emph{(B)} The multi-scale community organization of the synthetic dynamic network. Here we provide the results of the multi-scale dynamic community detection algorithm over several temporal scales. Note that the results at the lower $\omega$ values uncover fast dynamics whereas the results at the higher $\omega$ values uncover slow dynamics. \emph{(C)} The community allegiance matrices of the multi-scale communities over low, medium, and high $\omega$ values. Note that at higher temporal scales, the community allegiance matrices are similar to the combination of community allegiance matrices at lower temporal scales.}
\label{fig:Figure4}
\end{figure*}

\section{Hierarchical community organization of synthetic graphs}

Here we provide synthetic examples of graphs where each node can be identified locally within a community at three different topological scales. Next we create variations in the hierarchical organization of the graphs by systematically introducing edge strength inhomogeneities. As seen in Fig.~\ref{fig:Figure3}, multi-scale communities and the relative stability of communities across scales clearly uncovers the planted relationship (as well as the inhomogeneity profile) across the nodes.

\begin{figure*}
\centering
\setcounter{figure}{0}    
\includegraphics[width=.85\linewidth]{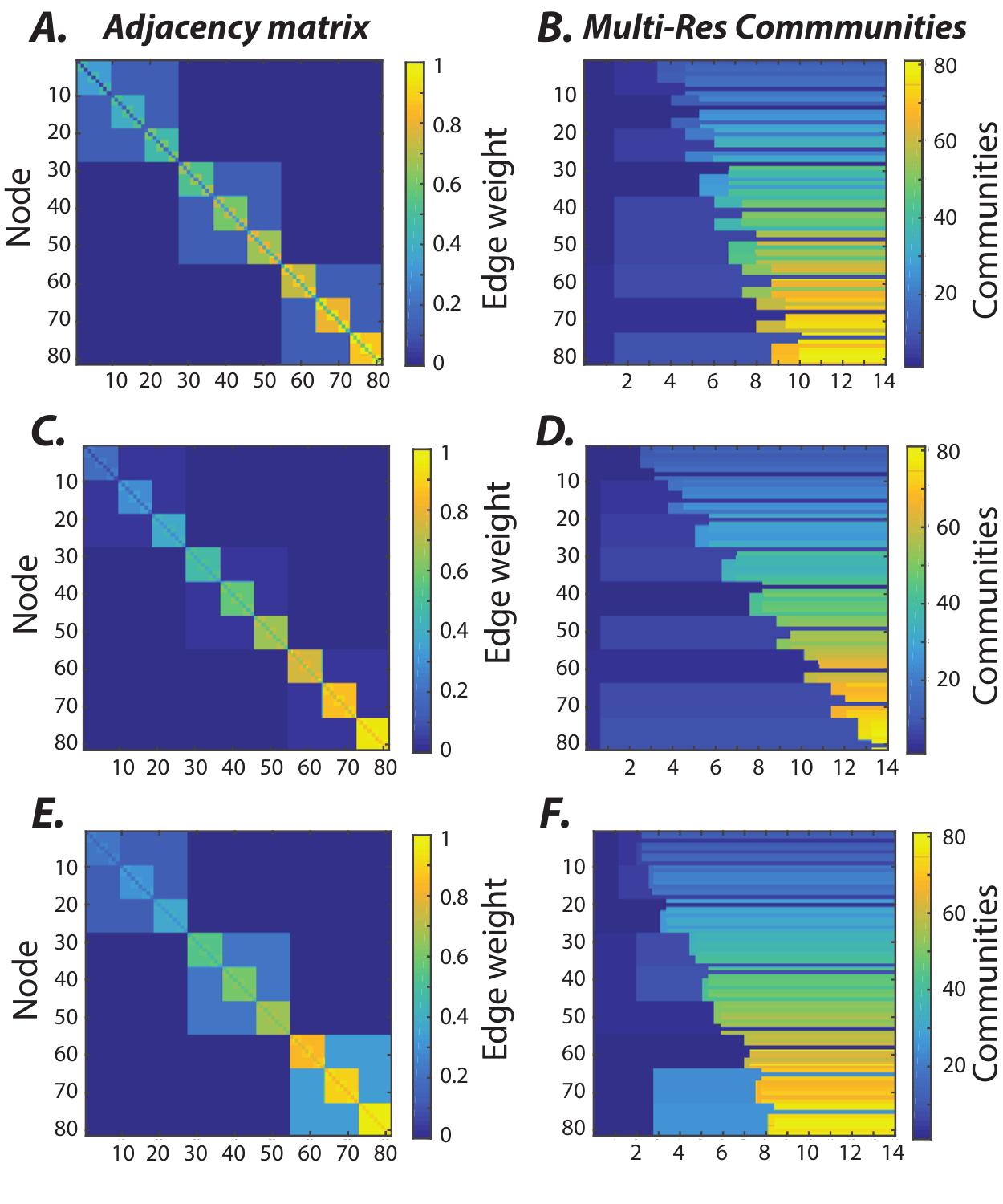}
\caption{\textbf{Synthetic graphs with diverse hierarchical structure.} The adjacency matrices (\emph{Left}) and their corresponding hierarchical community structure identified using the multi-scale method (\emph{Right}) are presented for three different graphs (A,C,E) where nodes form hierarchical communities at three separate topological scales. In graph (\emph{A}), the communities at coarse (and fine) topological scales are similarly identifiable across all nodes. However, the medium scale communities show differential discriminability as we have introduced node strength heterogeneity by assigning different within-community edge values for each community at that scale. The discriminability of these communities echoes the stability of the community branches: where stability is defined as the number of layers per branch divided by the total number of layers. The graph in (\emph{C}) displays the same three-scale topology as the graph in (\emph{A}) except that the medium scale communities have relatively higher discriminability, due to higher within-community edge weight at this scale. The higher stability of the branches of the multi-scale community (\emph{D}) similarly echoes the observation in (\emph{C})). The example in (\emph{E}) displays a graph where we introduced node strength heterogeneity in coarse- and medium-scale communities. Although the communities do not align at global scales for all nodes (as a result of the topological heterogeneity), the multi-scale community structure accurately preserves the local stability profiles of branches (\emph{F}), effectively revealing the three local topological scales in the graph. } 
\label{fig:Figure3}
\end{figure*}

\newpage
\section{Principal components analysis of the SC and FC stability matrices}

Here we use principal components analysis (PCA) to assess the stability profiles of nodes across $\gamma$ increments and measure the topological heterogeneity of the graphs. We used the number of components that account for more than $95\%$ of the variance in the stability matrices as a proxy for topological heterogeneity. Low numbers indicate that most nodes display similar stability profiles and therefore the graph is relatively topologically homogeneous. Conversely, higher numbers indicate that most nodes display diverse stability profiles and therefore the graph is relatively topologically heterogeneous (Fig.~\ref{fig:ResFigure3}).

\begin{figure*}[tbhp]
\centering
\setcounter{figure}{0}    
\includegraphics[width=.9\linewidth]{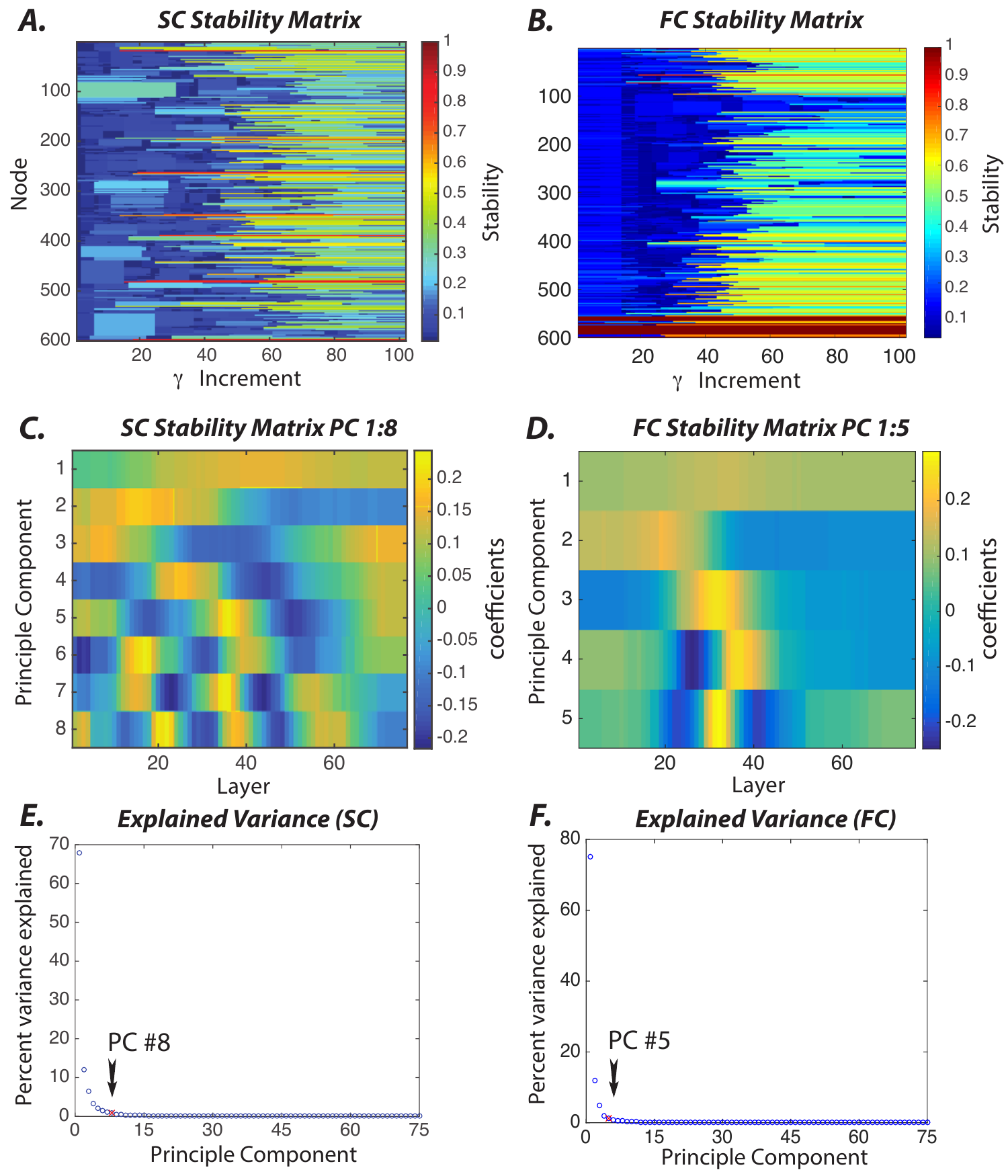}
\caption{\textbf{Principal components analysis of the SC and FC stability matrices extracted from a single representative subject}. \emph{(A)} The SC stability matrix and \emph{(B)} FC stability matrix created by exchanging the community labels with their corresponding stability values. \emph{(C$\&$D)} Principal components analysis of the matrices shown in panels \emph{(A$\&$B)}, with eight principal components shown for the SC stability matrix and five principal components shown for the FC stability matrix.  \emph{(E$\&$F)} We observe that only a small number of principal components account for most of the variance: eight components in the SC stability matrix and five components in the FC stability matrix account for $95 \%$ of the variance in the SC and FC stability matrices. Together, these results demonstrate that SC graphs are topologically more heterogeneous than FC graphs.}
\label{fig:ResFigure3}
\end{figure*}

\section{Hierarchical community organization of a multiplex SC and FC graph}

The FC and SC communities share similar community organization, and joint-optimization of FC and SC graphs (i.e. SC-FC multiplex graphs) can in theory be used to evaluate these similarities. That said, the community organization of the SC-FC multiplex graph is highly dependent on the inter-modality coupling parameter, $\kappa$. In Fig.~\ref{fig:ResFigure6}, we demonstrate that at smaller $\kappa$ values the FC-SC graph yields two separate community structures for FC and SC components of the graph; however, for higher $\kappa$ values they both share the same hybrid hierarchical community structure. The direct comparison between the community allegiance matrices of the FC, SC, and SC-FC graphs provided in Fig.~\ref{fig:ResFigure7} shows the effect of increasing inter-modality coupling parameter on the SC-FC community structure.

\begin{figure*}
\centering
\setcounter{figure}{0}    
\includegraphics[width=0.9\linewidth]{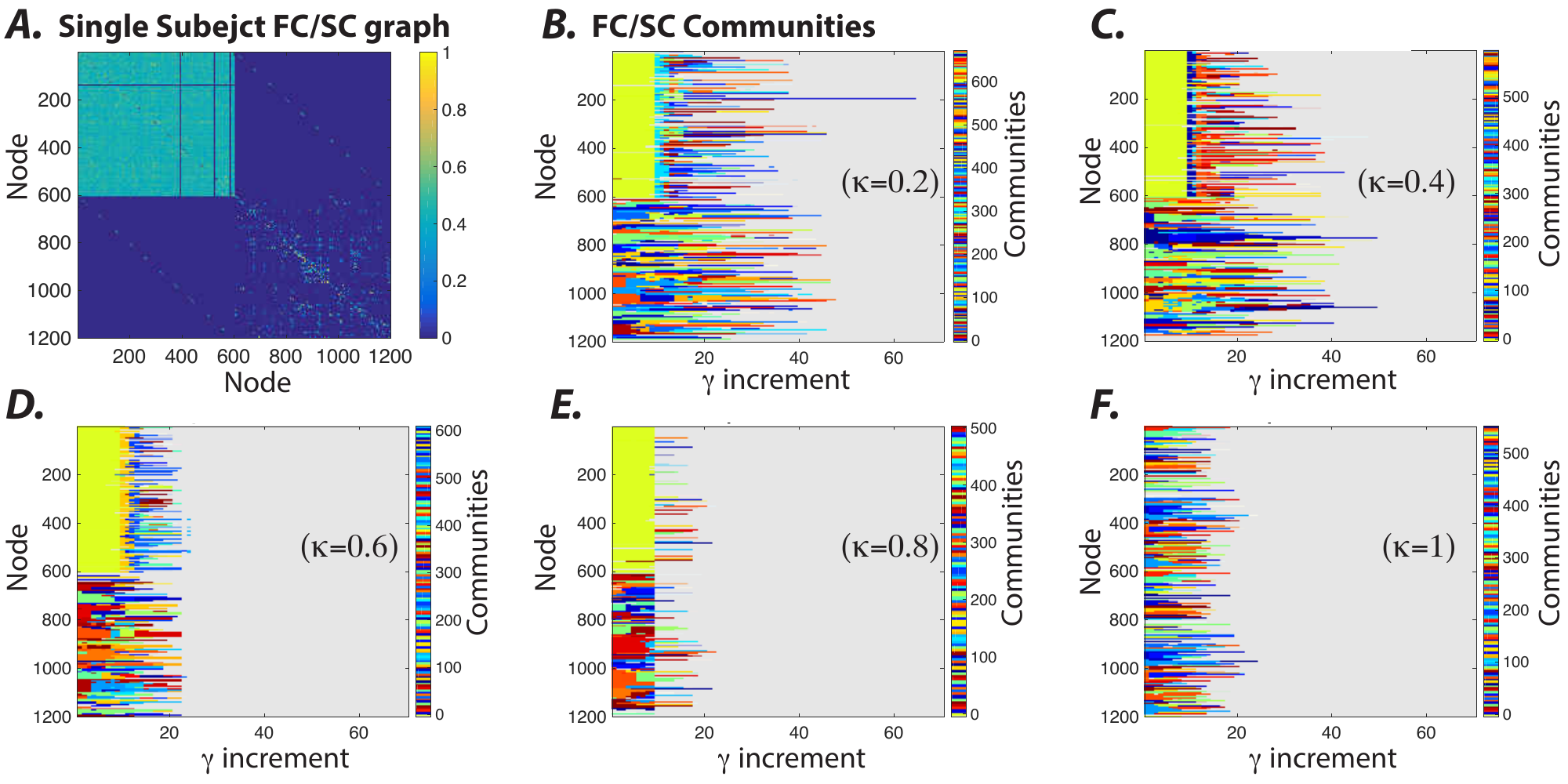}
\caption{\textbf{Hierarchical community organization of a multiplex, multi-scale graph that explicitly combines both imaging modalities: structure and function.}  \emph{(A)} The multiplex graph representing both imaging modalities is produced by coupling the nodes with the same identity across modalities (i.e., across FC and SC graphs). The strength of the coupling between modalities, $\kappa$, affects the dependence of the hierarchical community structure on both modalities. As $\kappa$ is tuned down, community structure is allowed to be independent in the SC and FC graphs, while when $\kappa$ is tuned up, community structure is forced to be more consistent across the two types of graphs. Panels \emph{(B-F)} demonstrate how increasing $\kappa$ causes a shift from two different hierarchical community structures for the FC and SC components of the multiplex graph \emph{(B)} to a single hybrid hierarchical community structure shared by both FC and SC components of the multiplex graph \emph{(F)}. } 
\label{fig:ResFigure6}
\end{figure*}

\begin{figure*}
\centering
\includegraphics[width=0.9\linewidth]{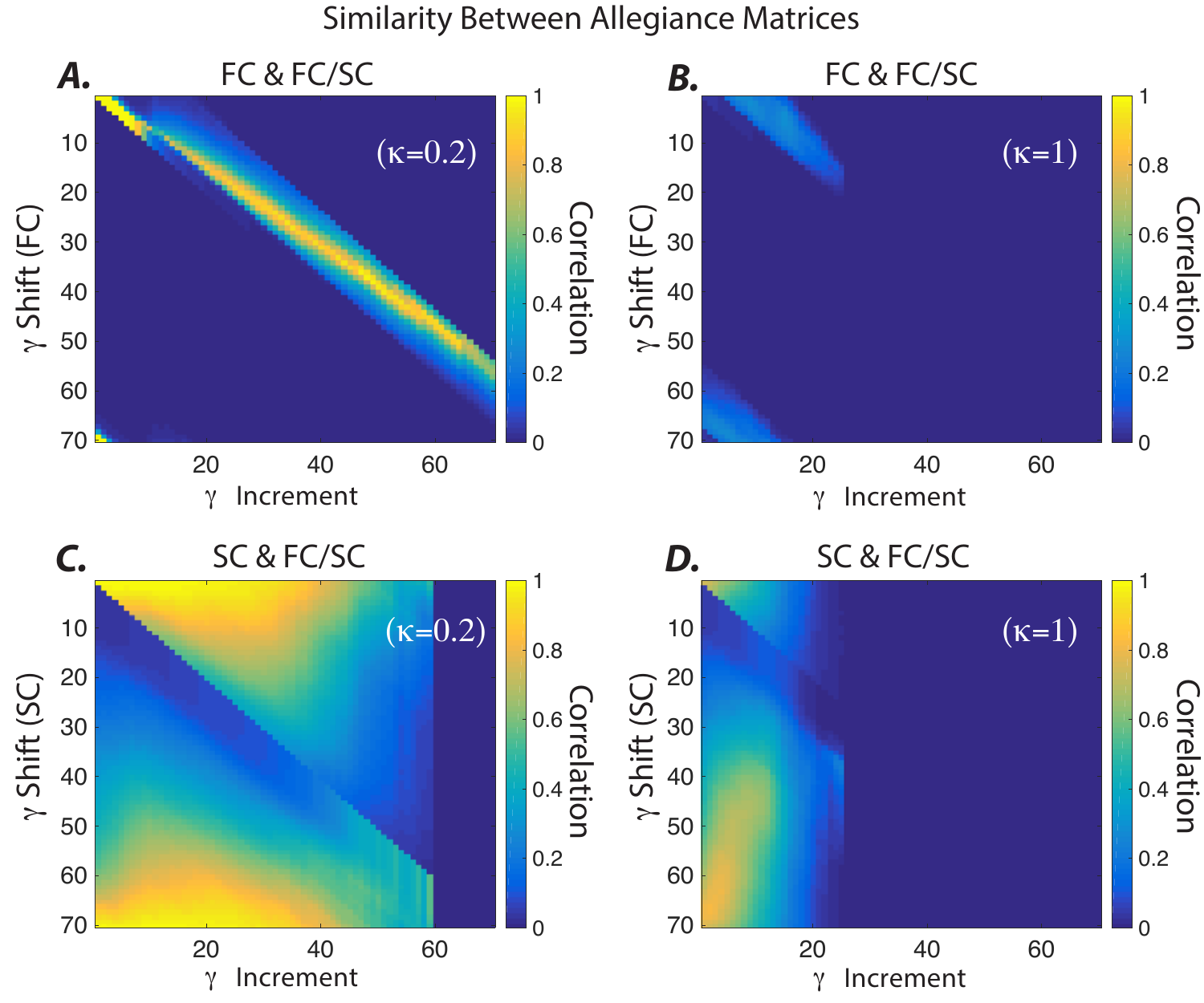} \caption{\textbf{Similarity between the allegiance matrices of the multiplex SC-FC graph, the FC graph alone, and the SC graph alone, as a function of the topological scale for a representative subject}. \emph{(A-B)} Pearson correlation coefficient values for all layers and all $\gamma$ shifts between the allegiance matrices of the multiplex SC-FC graph and the allegiance matrices of the FC graph alone, for low $(\kappa = 0.2)$ and high $(\kappa = 1)$ coupling values, respectively. Panels \emph{(C-D)} Pearson correlation coefficient values for all layers and all $\gamma$ shifts between the allegiance matrices of the multiplex SC-FC graph and the allegiance matrices of the SC graph alone, for low $(\kappa = 0.2)$ and high $(\kappa = 1)$ coupling values, respectively. }
\label{fig:ResFigure7}
\end{figure*}

\section{Sorting nodes based on multi-scale community allegiance}

A note on visualization. Sorting the nodes of the adjacency matrices based on their community allegiance allows us to visualize communities of densely connected nodes. Nevertheless for the multi-scale communities, the order of the nodes can change depending on the topological scale. In an effort to by-pass this limitation and enhance the visualization of these communities we sort the nodes based on the similarity of their community assignments across scales. Specifically, we perform optimal leaf ordering (optimalleaforder.m) for hierarchical clustering (linkage.m) using the distances (pdist.m) calculated between the community assignments of each pair of nodes. All multi-scale community plots in this manuscript were generated using this node sorting algorithm. 

\section{Multi-scale group consensus communities in structural and functional brain graphs}

Here we provide the group consensus multi-scale community results for the structural (Fig.~\ref{fig:SI_GCSC}) and functional graphs (Fig.~\ref{fig:SI_GSCFC}). One salient feature of the group consensus SC multi-scale community is that the communities are overwhelmingly made up of neighboring brain structures across the entire range of topological scales. To highlight the spatial proximity of the communities of the structural connectivity graphs, we identified (bwconncomp.m) and only presented the communities with more than one cluster in Fig.~\ref{fig:SI_GCSC_dis} while removing all the other communities with only one cluster of brain regions. Next, we tested the statistical significance of these observations via permutation test ($N=1000$) across $\gamma$ increments Fig.~\ref{fig:NumDiscon}. Our results demonstrate that the number of SC communities with more than one cluster is significantly ($p <0.01$, Bonferroni corrected for multiple comparisons) smaller than that of the null distribution (generated by changing the assignment of nodes to communities uniformly at random) for all the increments of $\gamma$  (except $\gamma = 3 $). Unlike structural graphs, functional graphs fail to yield group level consensus results for a large number brain regions, including several subcortical and cortical structures (Fig.~\ref{fig:SI_GSCFC}). We highlighted these structures separately in Fig.~\ref{fig:SI_GSCFC_single}.

\begin{figure*}[tbhp]
\centering
\setcounter{figure}{0}    
\includegraphics[width=1\linewidth]{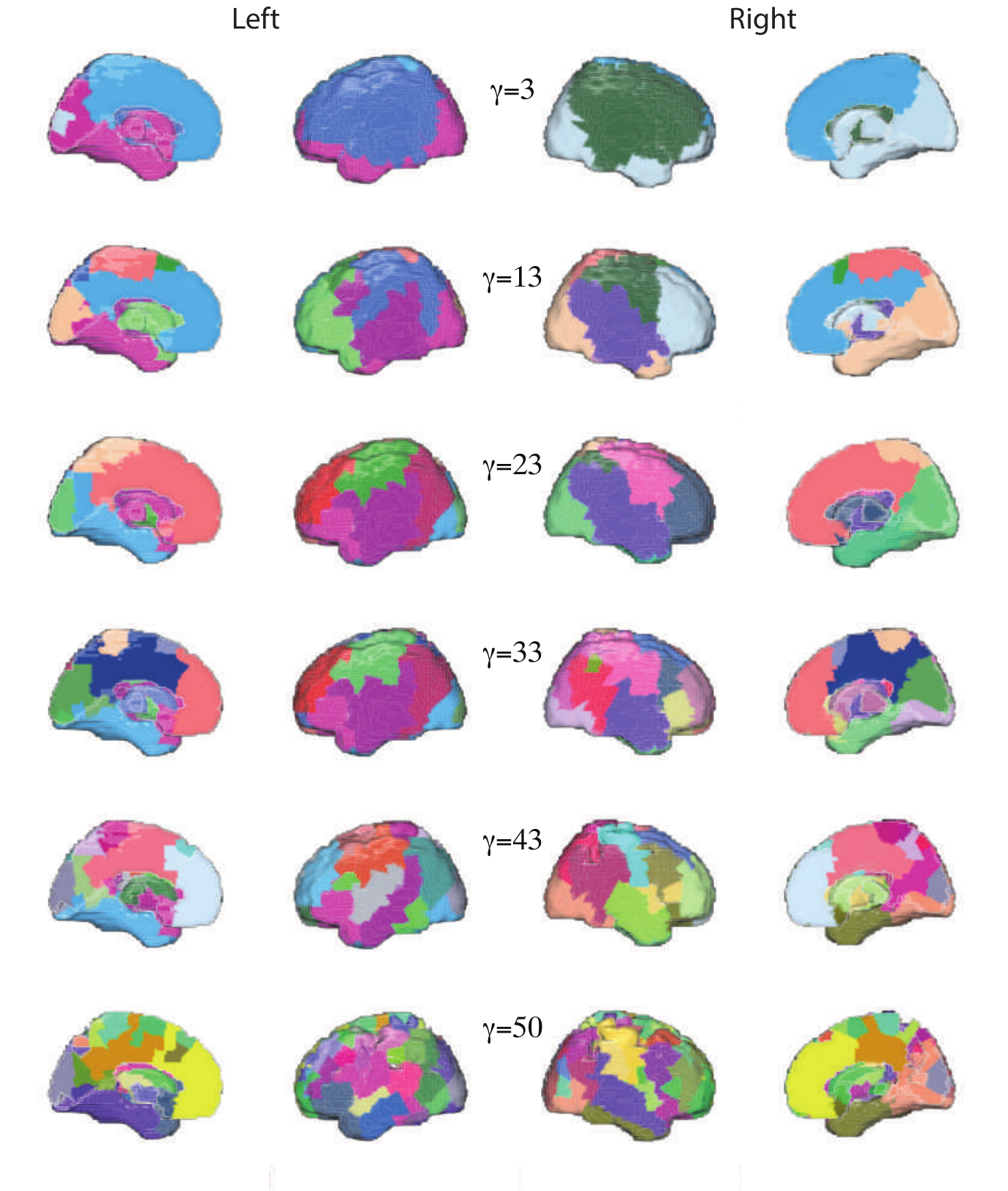}
\caption{\textbf{Group consensus hierarchical community organization of structural graphs}. The communities at several $\gamma$ values are color-coded and overlaid separately for each scale. Singleton communities are removed from the visualization for clarity.}
\label{fig:SI_GCSC}
\end{figure*}

\begin{figure*}[tbhp]
\centering
\includegraphics[width=1\linewidth]{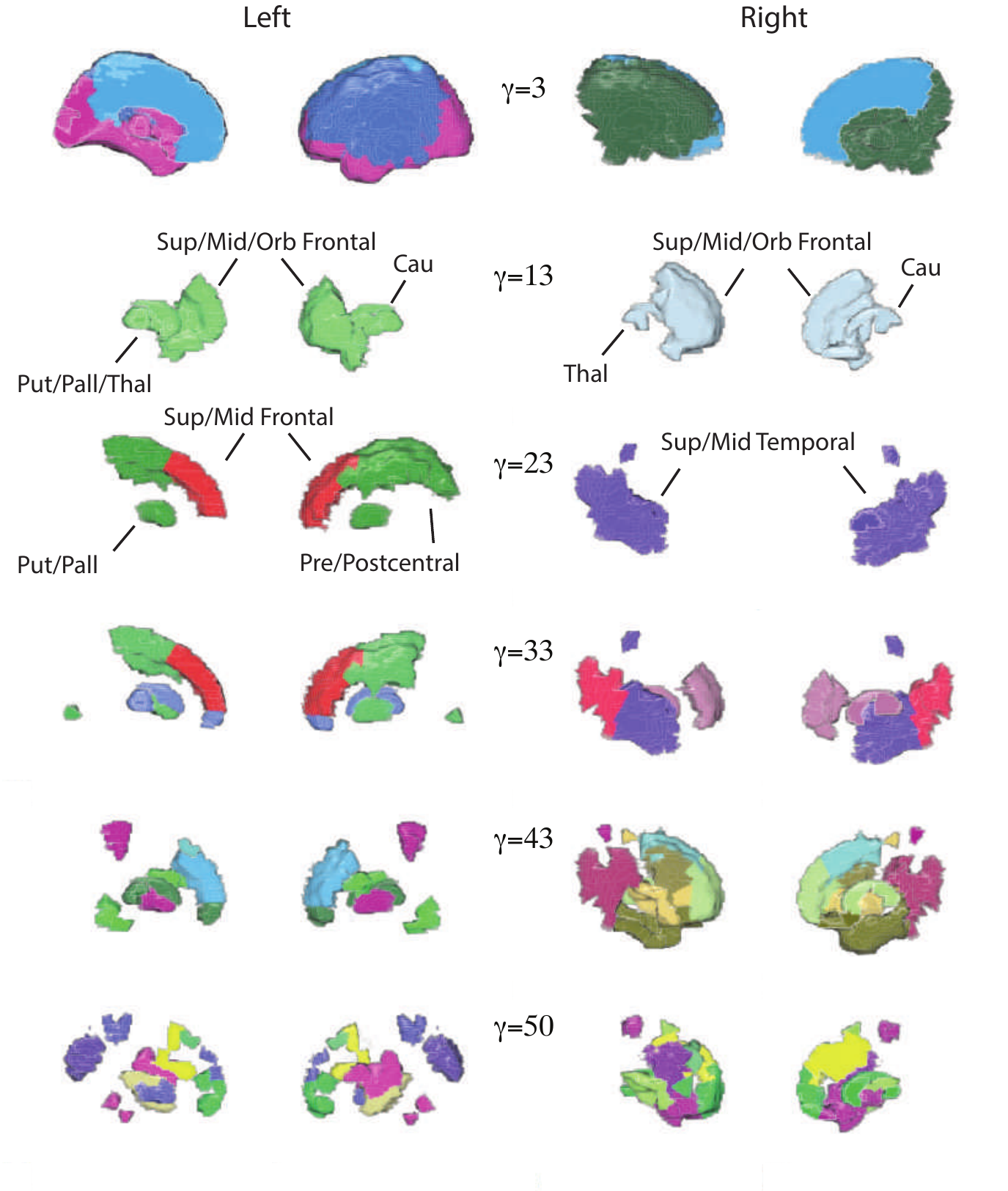}
\caption{\textbf{Spatially disconnected communities of the group-level hierarchical community structure in structural brain graphs}. Only the spatially disconnected communities (i.e., multi-cluster communities) at several $\gamma$ values are color-coded and overlaid separately for each scale (for details, see Appendix F). Singleton communities are removed. Note that several subcortical structures such as the caudate nucleus (Cau), pallidum (Pall), putamen (Put), and thalamus (Thal) form communities with areas in fronto-temporal cortex.}
\label{fig:SI_GCSC_dis}
\end{figure*}

\begin{figure*}[tbhp]
\centering
\includegraphics[width=1\linewidth]{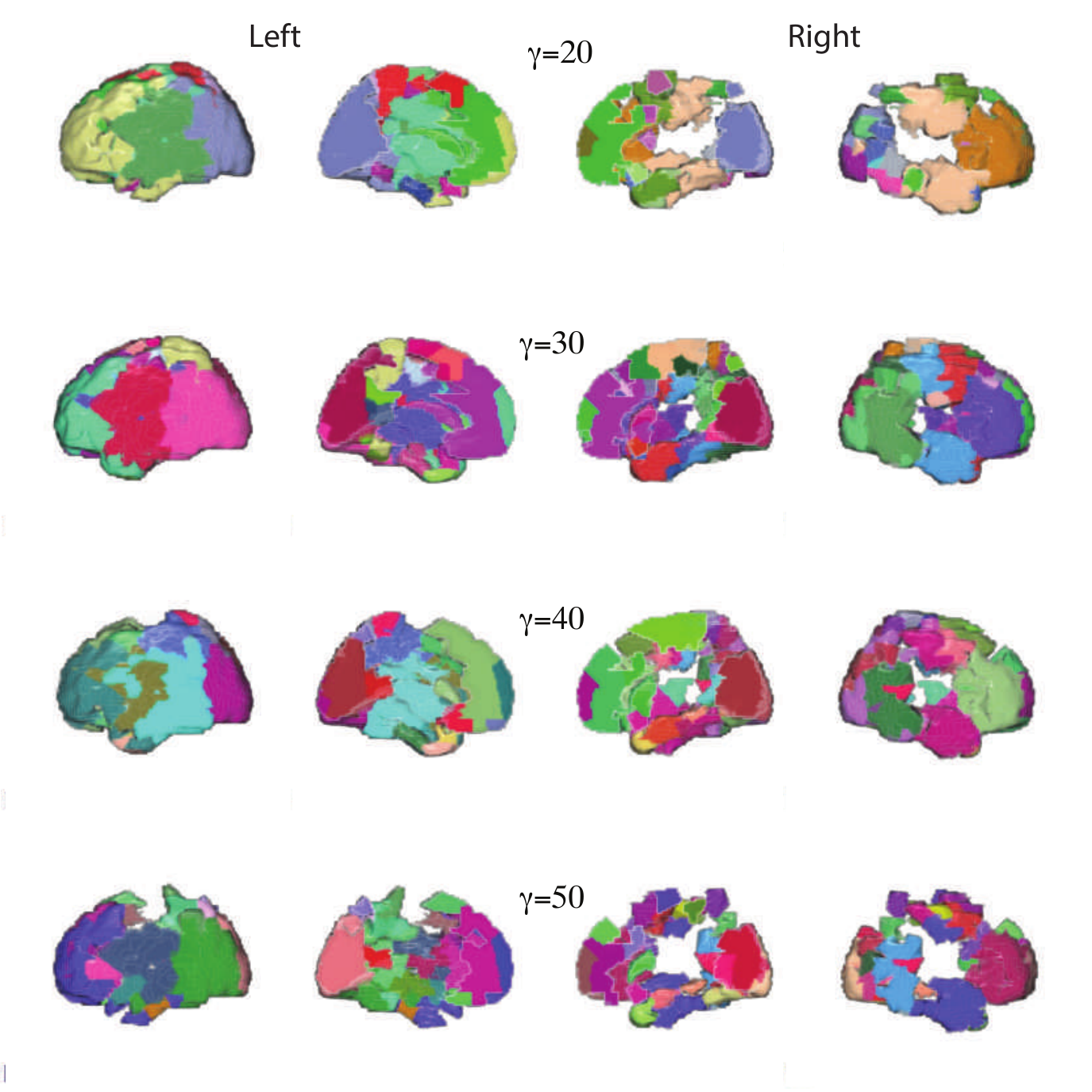}
\caption{\textbf{Group consensus hierarchical community organization of functional graphs}. The communities at several $\gamma$ values are color-coded and overlaid separately for each scale. Singleton communities are removed from the visualization for clarity.}
\label{fig:SI_GSCFC}
\end{figure*}

\begin{figure*}[tbhp]
\centering
\includegraphics[width=1\linewidth]{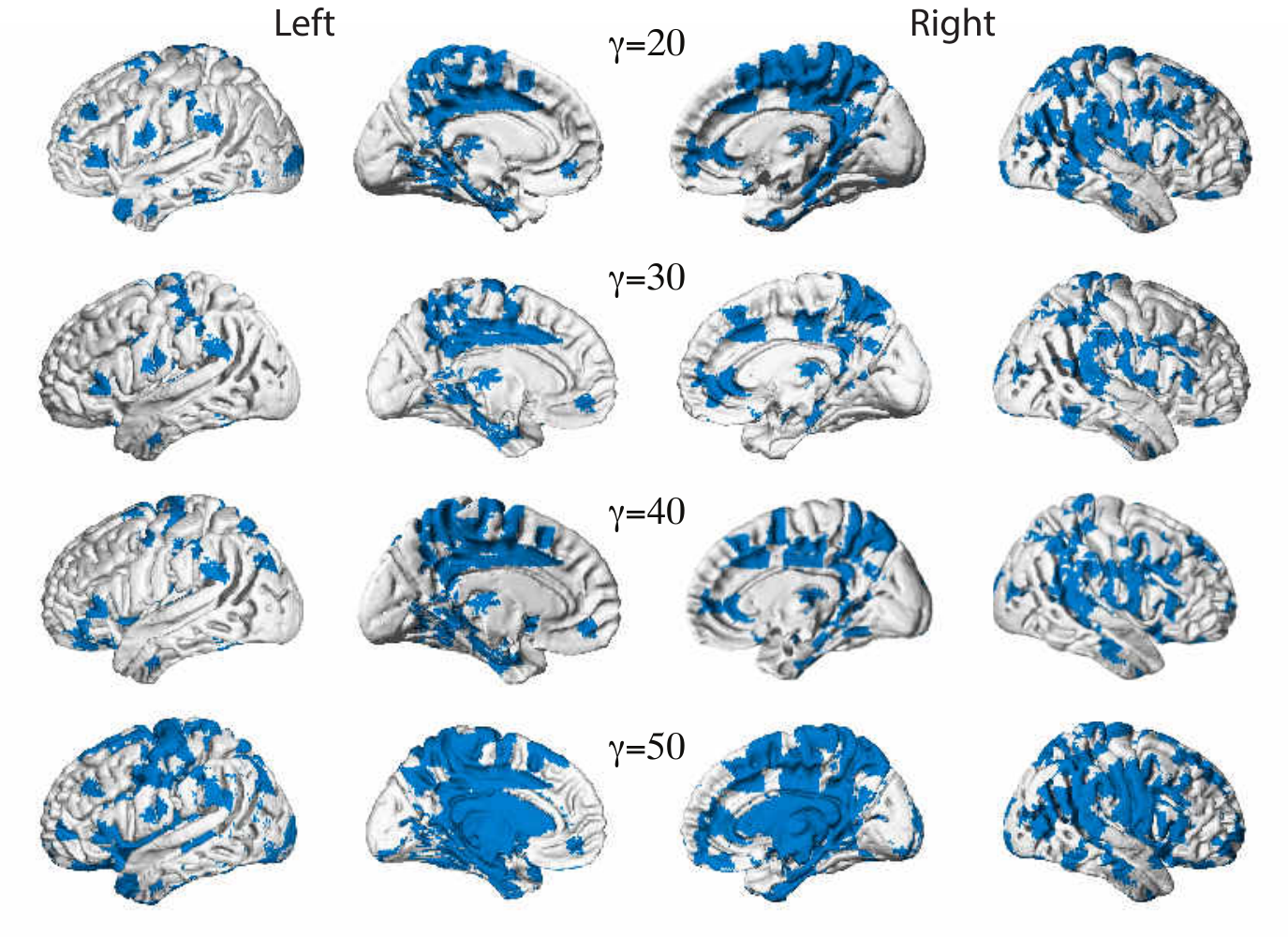}
\caption{\textbf{Regions lacking group-level consensus in functional connectivity graphs' hierarchical community structure}. Brain regions that fail to show significant group-level allegiance with other regions are identified as singleton communities in the consensus communities. The community allegiance of a pair of nodes is deemed significant if the group-level average allegiance of that pair (constructed from the subject-level consensus partitions) exceeds that of the null distribution constructed via randomizing the community label assignments.  All singleton communities are color-coded (blue) and overlaid on the cortical surface separately for different $\gamma$ values. These regions include subcortical structures (e.g., basal ganglia and thalamus) and large sections of posterior cingulate and dorsolateral (pre)frontal cortex. Note that these regions are identified as singletons even at coarse topological scales. }
\label{fig:SI_GSCFC_single}
\end{figure*}

\begin{figure*}[tbhp]
\centering
\includegraphics[width=1\linewidth]{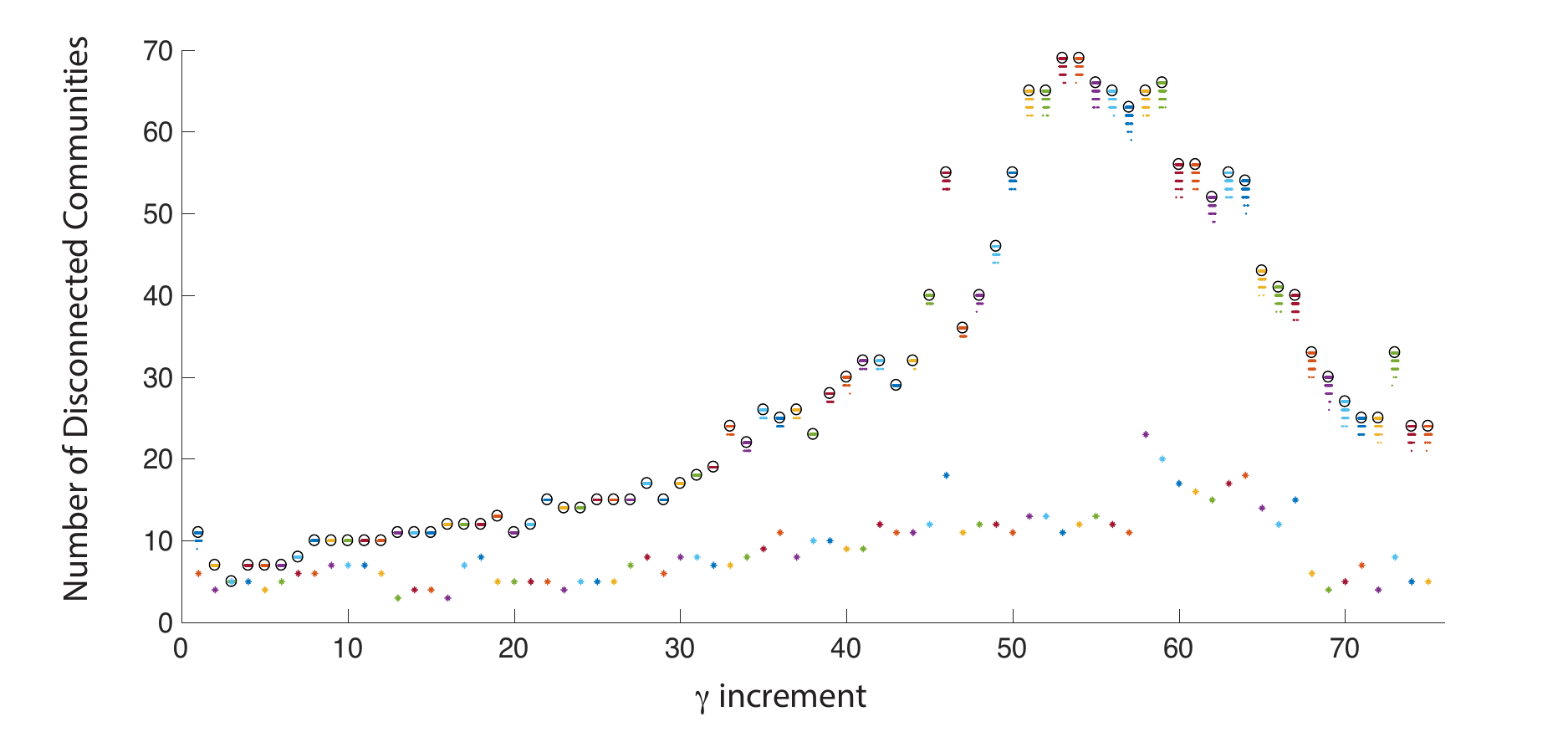}
\caption{\textbf{Number of spatially disconnected communities of the group-level hierarchical community structure in structural brain graphs}. The total number of communities and the total number of spatially disconnected communities (i.e., multi-cluster communities) at each $\gamma$ value are marked by \emph{'o'} and \emph{'*'}, respectively. The null distributions ($N=1000$) of the total number of spatially disconnected communities (created by randomizing the community labels) are marked by \emph{'.'} and presented for all $\gamma$ increments. Note that -- except for $\gamma = 3 $ -- all the empirical values are significantly lower than that of the null distribution. These results suggest that except for a few distributed networks, the SC communities at low $\gamma$ values mainly consist of a large number of neighboring nodes and branch into smaller communities of neighboring nodes. }
\label{fig:NumDiscon}
\end{figure*}

\section{Distribution of edge weights in the structural and functional brain graphs}

The distribution of the edges in the structural and functional connectivity matrices are notably different. While the distribution of SC edges are extremely heavy-tailed, the FC edges are close to a Gaussian distribution  (Fig.~\ref{fig:SI_Fig_edgedist}).

\begin{figure*}[tbhp]
\centering
\setcounter{figure}{0}    
\includegraphics[width=1\linewidth]{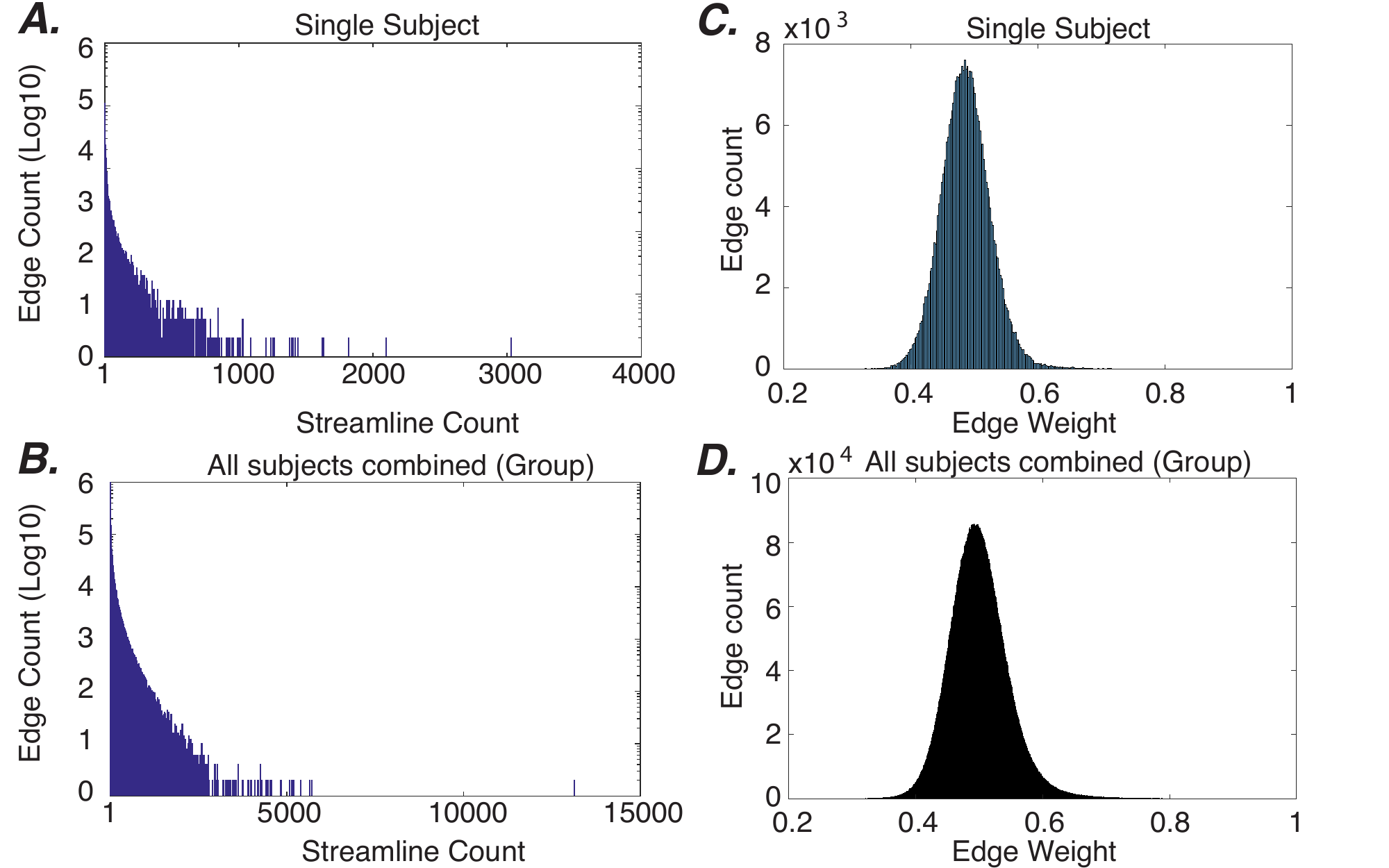}
\caption{\textbf{Distribution of edge weights in the structural and functional brain graphs.} The distribution of streamline counts for all edges in the structural brain graphs at the subject level \emph{(A)} and at the group level \emph{(B)}. Note the heavy-tailed distribution of streamline counts where hundreds of edges are made up of only a few streamlines while only a handful of edges are made up of hundreds of streamlines. Conversely, the functional brain graphs both at the subject level \emph{(C)} and at the group level \emph{(D)}) display a sharp Gaussian distribution where the vast majority of edges share similar values around the mean. }
\label{fig:SI_Fig_edgedist}
\end{figure*}

\pagebreak



  \bibliographystyle{elsarticle-harv} 
    \bibliography{Multi_scale,Multi_scale_DB}

\begin{thebibliography}{126}
\expandafter\ifx\csname natexlab\endcsname\relax\def\natexlab#1{#1}\fi
\expandafter\ifx\csname url\endcsname\relax
  \def\url#1{\texttt{#1}}\fi
\expandafter\ifx\csname urlprefix\endcsname\relax\def\urlprefix{URL }\fi

\bibitem[{Achard et~al.(2008)Achard, Bassett, Meyer-Lindenberg, and
  Bullmore}]{achard2008fractal}
Achard, S., Bassett, D.~S., Meyer-Lindenberg, A., Bullmore, E., 2008. Fractal
  connectivity of long-memory networks. Phys Rev E Stat Nonlin Soft Matter Phys
  77~(3 Pt 2), 036104.

\bibitem[{Akyildiz et~al.(2005)Akyildiz, Wang, and Wang}]{akyildiz2005wireless}
Akyildiz, I.~F., Wang, X., Wang, W., 2005. Wireless mesh networks: a survey.
  Computer networks 47~(4), 445--487.

\bibitem[{Arcila et~al.(2014)Arcila, Betizeau, Cambronne, Guzman, Doerflinger,
  Bouhallier, Zhou, Wu, Rani, Bassett, Borello, Huissoud, Goodman, Dehay, and
  Kosik}]{arcila2014novel}
Arcila, M.~L., Betizeau, M., Cambronne, X.~A., Guzman, E., Doerflinger, N.,
  Bouhallier, F., Zhou, H., Wu, B., Rani, N., Bassett, D.~S., Borello, U.,
  Huissoud, C., Goodman, R.~H., Dehay, C., Kosik, K.~S., 2014. Novel primate
  {miRNAs} coevolved with ancient target genes in germinal zone-specific
  expression patterns. Neuron 81~(6), 1255--1262.

\bibitem[{Arenas et~al.(2006)Arenas, D{\'\i}az-Guilera, and
  P{\'e}rez-Vicente}]{arenas2006synchronization}
Arenas, A., D{\'\i}az-Guilera, A., P{\'e}rez-Vicente, C.~J., 2006.
  Synchronization reveals topological scales in complex networks. Physical
  review letters 96~(11), 114102.

\bibitem[{Arenas et~al.(2008)Arenas, Fernandez, and Gomez}]{arenas2008analysis}
Arenas, A., Fernandez, A., Gomez, S., 2008. Analysis of the structure of
  complex networks at different resolution levels. New Journal of Physics
  10~(5), 053039.

\bibitem[{Ashburner et~al.(1999)Ashburner, Friston,
  et~al.}]{ashburner1999nonlinear}
Ashburner, J., Friston, K.~J., et~al., 1999. Nonlinear spatial normalization
  using basis functions. Human brain mapping 7~(4), 254--266.

\bibitem[{Barabasi and Oltvai(2004)}]{barabasi2004network}
Barabasi, A.-L., Oltvai, Z.~N., 2004. Network biology: understanding the cell's
  functional organization. Nature reviews genetics 5~(2), 101--113.

\bibitem[{Bassett et~al.(2011{\natexlab{a}})Bassett, Brown, Deshpande, Carlson,
  and Grafton}]{bassett2011conserved}
Bassett, D.~S., Brown, J.~A., Deshpande, V., Carlson, J.~M., Grafton, S.~T.,
  2011{\natexlab{a}}. Conserved and variable architecture of human white matter
  connectivity. Neuroimage 54~(2), 1262--1279.

\bibitem[{Bassett et~al.(2010)Bassett, Greenfield, Meyer-Lindenberg,
  Weinberger, Moore, and Bullmore}]{bassett2010efficient}
Bassett, D.~S., Greenfield, D.~L., Meyer-Lindenberg, A., Weinberger, D.~R.,
  Moore, S.~W., Bullmore, E.~T., 2010. Efficient physical embedding of
  topologically complex information processing networks in brains and computer
  circuits. PLoS Comput Biol 6~(4), e1000748.

\bibitem[{Bassett et~al.(2015{\natexlab{a}})Bassett, Owens, Porter, Manning,
  and Daniels}]{bassett2015extraction}
Bassett, D.~S., Owens, E.~T., Porter, M.~A., Manning, M.~L., Daniels, K.~E.,
  2015{\natexlab{a}}. Extraction of force-chain network architecture in
  granular materials using community detection. Soft Matter 11~(14),
  2731--2744.

\bibitem[{Bassett et~al.(2013{\natexlab{a}})Bassett, Porter, Wymbs, Grafton,
  Carlson, and Mucha}]{bassett2013robust}
Bassett, D.~S., Porter, M.~A., Wymbs, N.~F., Grafton, S.~T., Carlson, J.~M.,
  Mucha, P.~J., 2013{\natexlab{a}}. Robust detection of dynamic community
  structure in networks. Chaos: An Interdisciplinary Journal of Nonlinear
  Science 23~(1), 013142.

\bibitem[{Bassett and Siebenhuhner(2013)}]{bassett2013multi}
Bassett, D.~S., Siebenhuhner, F., 2013. Multiscale network organization in the
  human brain. Wiley.

\bibitem[{Bassett et~al.(2011{\natexlab{b}})Bassett, Wymbs, Porter, Mucha,
  Carlson, and Grafton}]{bassett2011}
Bassett, D.~S., Wymbs, N.~F., Porter, M.~A., Mucha, P.~J., Carlson, J.~M.,
  Grafton, S.~T., 2011{\natexlab{b}}. Dynamic reconfiguration of human brain
  networks during learning. Proc Natl Acad Sci U S A 108~(18), 7641--7646.

\bibitem[{Bassett et~al.(2014)Bassett, Wymbs, Porter, Mucha, and
  Grafton}]{bassett2014cross}
Bassett, D.~S., Wymbs, N.~F., Porter, M.~A., Mucha, P.~J., Grafton, S.~T.,
  2014. Cross-linked structure of network evolution. Chaos 24~(1), 013112.

\bibitem[{Bassett et~al.(2013{\natexlab{b}})Bassett, Wymbs, Rombach, Porter,
  Mucha, and Grafton}]{bassett2013task}
Bassett, D.~S., Wymbs, N.~F., Rombach, M.~P., Porter, M.~A., Mucha, P.~J.,
  Grafton, S.~T., 2013{\natexlab{b}}. Task-based core-periphery organization of
  human brain dynamics. PLoS Comput Biol 9~(9), e1003171.

\bibitem[{Bassett et~al.(2015{\natexlab{b}})Bassett, Yang, Wymbs, and
  Grafton}]{bassett2015learning}
Bassett, D.~S., Yang, M., Wymbs, N.~F., Grafton, S.~T., 2015{\natexlab{b}}.
  Learning-induced autonomy of sensorimotor systems. Nat Neurosci 18~(5),
  744--751.

\bibitem[{Becker et~al.(2015)Becker, Pequito, Pappas, Miller, Grafton, Bassett,
  and Preciado}]{becker2015accurately}
Becker, C.~O., Pequito, S., Pappas, G.~J., Miller, M.~B., Grafton, S.~T.,
  Bassett, D.~S., Preciado, V.~M., 2015. Accurately predicting functional
  connectivity from diffusion imaging. arXiv preprint arXiv:1512.02602.

\bibitem[{Beggs(2008)}]{beggs2008criticality}
Beggs, J.~M., 2008. The criticality hypothesis: how local cortical networks
  might optimize information processing. Philosophical Transactions of the
  Royal Society of London A: Mathematical, Physical and Engineering Sciences
  366~(1864), 329--343.

\bibitem[{Berry et~al.(2011)Berry, Hendrickson, LaViolette, and
  Phillips}]{berry2011tolerating}
Berry, J.~W., Hendrickson, B., LaViolette, R.~A., Phillips, C.~A., 2011.
  Tolerating the community detection resolution limit with edge weighting. Phys
  Rev E Stat Nonlin Soft Matter Phys 83~(5 Pt 2), 056119.

\bibitem[{Betzel and Bassett(2016)}]{betzel2016multi}
Betzel, R.~F., Bassett, D.~S., 2016. Multi-scale brain networks. NeuroImage.

\bibitem[{Betzel et~al.(2013)Betzel, Griffa, Avena-Koenigsberger, Go{\~n}i,
  Thiran, Hagmann, and Sporns}]{betzel2013multi}
Betzel, R.~F., Griffa, A., Avena-Koenigsberger, A., Go{\~n}i, J., Thiran,
  J.-P., Hagmann, P., Sporns, O., 2013. Multi-scale community organization of
  the human structural connectome and its relationship with resting-state
  functional connectivity. Network Science 1~(03), 353--373.

\bibitem[{Betzel et~al.(2016)Betzel, Gu, Medaglia, Pasqualetti, and
  Bassett}]{betzel2016optimally}
Betzel, R.~F., Gu, S., Medaglia, J.~D., Pasqualetti, F., Bassett, D.~S., 2016.
  Optimally controlling the human connectome: the role of network topology. Sci
  Rep 6, 30770.

\bibitem[{Blondel et~al.(2008)Blondel, Guillaume, Lambiotte, and
  Lefebvre}]{blondel2008fast}
Blondel, V.~D., Guillaume, J.-L., Lambiotte, R., Lefebvre, E., 2008. Fast
  unfolding of communities in large networks. J. Stat. Mech., P10008.

\bibitem[{Braun et~al.(2015)Braun, Schafer, Walter, Erk, Romanczuk-Seiferth,
  Haddad, Schweiger, Grimm, Heinz, Tost, Meyer-Lindenberg, and
  Bassett}]{braun2015dynamic}
Braun, U., Schafer, A., Walter, H., Erk, S., Romanczuk-Seiferth, N., Haddad,
  L., Schweiger, J.~I., Grimm, O., Heinz, A., Tost, H., Meyer-Lindenberg, A.,
  Bassett, D.~S., 2015. Dynamic reconfiguration of frontal brain networks
  during executive cognition in humans. Proc Natl Acad Sci U S A 112~(37),
  11678--11683.

\bibitem[{Brun et~al.(2009)Brun, Lepor{\'e}, Pennec, Lee, Barysheva, Madsen,
  Avedissian, Chou, De~Zubicaray, McMahon, et~al.}]{brun2009mapping}
Brun, C.~C., Lepor{\'e}, N., Pennec, X., Lee, A.~D., Barysheva, M., Madsen,
  S.~K., Avedissian, C., Chou, Y.-Y., De~Zubicaray, G.~I., McMahon, K.~L.,
  et~al., 2009. Mapping the regional influence of genetics on brain structure
  variability?a tensor-based morphometry study. Neuroimage 48~(1), 37--49.

\bibitem[{Bullmore et~al.(2009)Bullmore, Barnes, Bassett, Fornito, Kitzbichler,
  Meunier, and Suckling}]{bullmore2009generic}
Bullmore, E., Barnes, A., Bassett, D.~S., Fornito, A., Kitzbichler, M.,
  Meunier, D., Suckling, J., 2009. Generic aspects of complexity in brain
  imaging data and other biological systems. Neuroimage 47~(3), 1125--1134.

\bibitem[{Bullmore and Sporns(2009)}]{bullmore2009complex}
Bullmore, E., Sporns, O., 2009. Complex brain networks: graph theoretical
  analysis of structural and functional systems. Nat Rev Neurosci 10~(3),
  186--198.

\bibitem[{Bullmore and Sporns(2012)}]{bullmore2012economy}
Bullmore, E., Sporns, O., 2012. The economy of brain network organization. Nat
  Rev Neurosci 13~(5), 336--349.

\bibitem[{Calhoun et~al.(2014)Calhoun, Miller, Pearlson, and
  Adali}]{calhoun2014chronnectome}
Calhoun, V.~D., Miller, R., Pearlson, G., Adali, T., 2014. The chronnectome:
  time-varying connectivity networks as the next frontier in {fMRI} data
  discovery. Neuron 84~(2), 262--274.

\bibitem[{Chai et~al.(2017)Chai, Khambhati, Ciric, Gur, Gur, Satterthwaite, and
  Bassett}]{chai2017evolution}
Chai, L., Khambhati, A.~N., Ciric, R., Gur, R.~C., Gur, R.~E., Satterthwaite,
  T.~D., Bassett, D.~S., 2017. Evolution of brain network dynamics in
  neurodevelopment. Network Neuroscience In Press.

\bibitem[{Chai et~al.(2016)Chai, Mattar, Blank, Fedorenko, and
  Bassett}]{chai2016functional}
Chai, L.~R., Mattar, M.~G., Blank, I.~A., Fedorenko, E., Bassett, D.~S., 2016.
  Functional network dynamics of the language system. Cereb Cortex Epub ahead
  of print.

\bibitem[{Chaudhuri et~al.(2014)Chaudhuri, Bernacchia, and
  Wang}]{chaudhuri2014diversity}
Chaudhuri, R., Bernacchia, A., Wang, X.~J., 2014. A diversity of localized
  timescales in network activity. Elife 3, e01239.

\bibitem[{Chen(2016)}]{chen2016vlsi}
Chen, W.-K., 2016. The VLSI handbook. CRC press.

\bibitem[{Chialvo et~al.(2008)Chialvo, Balenzuela, and
  Fraiman}]{chialvo2008brain}
Chialvo, D.~R., Balenzuela, P., Fraiman, D., 2008. The brain: what is critical
  about it? arXiv preprint arXiv:0804.0032.

\bibitem[{Ciric et~al.(2016)Ciric, Wolf, Power, Roalf, Baum, Ruparel,
  Shinohara, Elliott, Eickhoff, Davatzikos, Gur, Gur, Bassett, and
  Satterthwaite}]{rastko2016benchmarking}
Ciric, R., Wolf, D.~H., Power, J.~D., Roalf, D.~R., Baum, G., Ruparel, K.,
  Shinohara, R.~T., Elliott, M.~A., Eickhoff, S.~B., Davatzikos, C., Gur,
  R.~C., Gur, R.~E., Bassett, D.~S., Satterthwaite, T.~D., 2016. Benchmarking
  confound regression strategies for the control of motion artifact in studies
  of functional connectivity. arXiv 1608, 03616.

\bibitem[{Conaco et~al.(2012)Conaco, Bassett, Zhou, Arcila, Degnan, Degnan, and
  Kosik}]{conaco2012functionalization}
Conaco, C., Bassett, D.~S., Zhou, H., Arcila, M.~L., Degnan, S.~M., Degnan,
  B.~M., Kosik, K.~S., 2012. Functionalization of a protosynaptic gene
  expression network. Proc Natl Acad Sci U S A 109~(Suppl 1), 10612--10618.

\bibitem[{Croxson et~al.(2005)Croxson, Johansen-Berg, Behrens, Robson, Pinsk,
  Gross, Richter, Richter, Kastner, and Rushworth}]{croxson2005quantitative}
Croxson, P.~L., Johansen-Berg, H., Behrens, T.~E., Robson, M.~D., Pinsk, M.~A.,
  Gross, C.~G., Richter, W., Richter, M.~C., Kastner, S., Rushworth, M.~F.,
  2005. Quantitative investigation of connections of the prefrontal cortex in
  the human and macaque using probabilistic diffusion tractography. The Journal
  of neuroscience 25~(39), 8854--8866.

\bibitem[{Danon et~al.(2006)Danon, D{\'\i}az-Guilera, and
  Arenas}]{danon2006effect}
Danon, L., D{\'\i}az-Guilera, A., Arenas, A., 2006. The effect of size
  heterogeneity on community identification in complex networks. Journal of
  Statistical Mechanics: Theory and Experiment 2006~(11), P11010.

\bibitem[{Delvenne et~al.(2010)Delvenne, Yaliraki, and
  Barahona}]{delvenne2010stability}
Delvenne, J.-C., Yaliraki, S.~N., Barahona, M., 2010. Stability of graph
  communities across time scales. Proceedings of the National Academy of
  Sciences 107~(29), 12755--12760.

\bibitem[{Desikan et~al.(2006)Desikan, S{\'e}gonne, Fischl, Quinn, Dickerson,
  Blacker, Buckner, Dale, Maguire, Hyman, et~al.}]{desikan2006automated}
Desikan, R.~S., S{\'e}gonne, F., Fischl, B., Quinn, B.~T., Dickerson, B.~C.,
  Blacker, D., Buckner, R.~L., Dale, A.~M., Maguire, R.~P., Hyman, B.~T.,
  et~al., 2006. An automated labeling system for subdividing the human cerebral
  cortex on mri scans into gyral based regions of interest. Neuroimage 31~(3),
  968--980.

\bibitem[{Diez et~al.(2015)Diez, Bonifazi, Escudero, Mateos, Mu{\~n}oz,
  Stramaglia, and Cortes}]{diez2015novel}
Diez, I., Bonifazi, P., Escudero, I., Mateos, B., Mu{\~n}oz, M.~A., Stramaglia,
  S., Cortes, J.~M., 2015. A novel brain partition highlights the modular
  skeleton shared by structure and function. Scientific reports 5.

\bibitem[{Fenn et~al.(2009)Fenn, Porter, McDonald, Williams, Johnson, and
  Jones}]{fenn2009dynamic}
Fenn, D.~J., Porter, M.~A., McDonald, M., Williams, S., Johnson, N.~F., Jones,
  N.~S., 2009. Dynamic communities in multichannel data: An application to the
  foreign exchange market during the 2007--2008 credit crisis. Chaos: An
  Interdisciplinary Journal of Nonlinear Science 19~(3), 033119.

\bibitem[{Fenn et~al.(2012)Fenn, Porter, Mucha, McDonald, Williams, Johnson,
  and Jones}]{fenn2012dynamical}
Fenn, D.~J., Porter, M.~A., Mucha, P.~J., McDonald, M., Williams, S., Johnson,
  N.~F., Jones, N.~S., 2012. Dynamical clustering of exchange rates.
  Quantitative Finance 12~(10), 1493--1520.

\bibitem[{Finn et~al.(2015)Finn, Shen, Scheinost, Rosenberg, Huang, Chun,
  Papademetris, and Constable}]{finn2015functional}
Finn, E.~S., Shen, X., Scheinost, D., Rosenberg, M.~D., Huang, J., Chun, M.~M.,
  Papademetris, X., Constable, R.~T., 2015. Functional connectome
  fingerprinting: identifying individuals using patterns of brain connectivity.
  Nature neuroscience.

\bibitem[{Fortunato(2010)}]{fortunato2010}
Fortunato, S., 2010. Community detection in graphs. Physics reports 486~(3),
  75--174.

\bibitem[{Gerraty et~al.(2016)Gerraty, Davidow, Foerde, Galvan, Bassett, and
  Shohamy}]{gerraty2016dynamic}
Gerraty, R.~T., Davidow, J.~Y., Foerde, K., Galvan, A., Bassett, D.~S.,
  Shohamy, D., 2016. Dynamic flexibility in striatal-cortical circuits supports
  reinforcement learning. bioRxiv 094383~(1101), 094383.

\bibitem[{Girvan and Newman(2002)}]{girvan2002}
Girvan, M., Newman, M. E.~J., 2002. Community structure in social and
  biological networks. Proc. Natl. Acad. Sci. USA 99, 7821--7826.

\bibitem[{Giusti et~al.(2016)Giusti, Ghrist, and Bassett}]{giusti2016twos}
Giusti, C., Ghrist, R., Bassett, D.~S., 2016. Two's company, three (or more) is
  a simplex: {A}lgebraic-topological tools for understanding higher-order
  structure in neural data. J Comput Neurosci 41~(1), 1--14.

\bibitem[{Go{\~n}i et~al.(2014)Go{\~n}i, van~den Heuvel, Avena-Koenigsberger,
  de~Mendizabal, Betzel, Griffa, Hagmann, Corominas-Murtra, Thiran, and
  Sporns}]{goni2014resting}
Go{\~n}i, J., van~den Heuvel, M.~P., Avena-Koenigsberger, A., de~Mendizabal,
  N.~V., Betzel, R.~F., Griffa, A., Hagmann, P., Corominas-Murtra, B., Thiran,
  J.-P., Sporns, O., 2014. Resting-brain functional connectivity predicted by
  analytic measures of network communication. Proceedings of the National
  Academy of Sciences 111~(2), 833--838.

\bibitem[{Gonzalez-Castillo et~al.(2014)Gonzalez-Castillo, Handwerker,
  Robinson, Hoy, Buchanan, Saad, and Bandettini}]{gonzalez2014spatial}
Gonzalez-Castillo, J., Handwerker, D.~A., Robinson, M.~E., Hoy, C.~W.,
  Buchanan, L.~C., Saad, Z.~S., Bandettini, P.~A., 2014. The spatial structure
  of resting state connectivity stability on the scale of minutes. Frontiers in
  neuroscience 8, 138.

\bibitem[{Good et~al.(2010)Good, de~Montjoye, and
  Clauset}]{good2010performance}
Good, B.~H., de~Montjoye, Y.~A., Clauset, A., 2010. Performance of modularity
  maximization in practical contexts. Phys Rev E Stat Nonlin Soft Matter Phys
  81~(4 Pt 2), 046106.

\bibitem[{Grinsted et~al.(2004)Grinsted, Moore, and Jevrejeva}]{grinsted2004}
Grinsted, A., Moore, J.~C., Jevrejeva, S., 2004. Application of the cross
  wavelet transform and wavelet coherence to geophysical time series. Nonlin.
  Processes Geophys. 11, 561--566.

\bibitem[{Gu et~al.(2015)Gu, Pasqualetti, Cieslak, Telesford, Yu, Kahn,
  Medaglia, Vettel, Miller, Grafton, and Bassett}]{gu2015controllability}
Gu, S., Pasqualetti, F., Cieslak, M., Telesford, Q.~K., Yu, A.~B., Kahn, A.~E.,
  Medaglia, J.~D., Vettel, J.~M., Miller, M.~B., Grafton, S.~T., Bassett,
  D.~S., 2015. Controllability of structural brain networks. Nat Commun 6,
  8414.

\bibitem[{Guimera and Amaral(2005)}]{guimera2005functional}
Guimera, R., Amaral, L. A.~N., 2005. Functional cartography of complex
  metabolic networks. Nature 433~(7028), 895--900.

\bibitem[{Hagmann et~al.(2008)Hagmann, Cammoun, Gigandet, Meuli, Honey, Wedeen,
  and Sporns}]{hagmann2008mapping}
Hagmann, P., Cammoun, L., Gigandet, X., Meuli, R., Honey, C.~J., Wedeen, V.~J.,
  Sporns, O., 2008. Mapping the structural core of human cerebral cortex. PLoS
  Biol 6~(7), e159.

\bibitem[{Helbing et~al.(2006)Helbing, Ammoser, and
  K{\"u}hnert}]{helbing2006information}
Helbing, D., Ammoser, H., K{\"u}hnert, C., 2006. Information flows in
  hierarchical networks and the capability of organizations to successfully
  respond to failures, crises, and disasters. Physica A: Statistical Mechanics
  and its Applications 363~(1), 141--150.

\bibitem[{Henzler et~al.(2013)Henzler, Li, Dang, Arcila, Zhou, Liu, Chang,
  Bassett, Rana, and Kosik}]{henzler2013staged}
Henzler, C.~M., Li, Z., Dang, J., Arcila, M.~L., Zhou, H., Liu, J., Chang,
  K.~Y., Bassett, D.~S., Rana, T.~M., Kosik, K.~S., 2013. Staged {miRNA}
  re-regulation patterns during reprogramming. Genome Biol 14~(12), R149.

\bibitem[{Hermundstad et~al.(2013)Hermundstad, Bassett, Brown, Aminoff,
  Clewett, Freeman, Frithsen, Johnson, Tipper, Miller,
  et~al.}]{hermundstad2013structural}
Hermundstad, A.~M., Bassett, D.~S., Brown, K.~S., Aminoff, E.~M., Clewett, D.,
  Freeman, S., Frithsen, A., Johnson, A., Tipper, C.~M., Miller, M.~B., et~al.,
  2013. Structural foundations of resting-state and task-based functional
  connectivity in the human brain. Proceedings of the National Academy of
  Sciences 110~(15), 6169--6174.

\bibitem[{Hermundstad et~al.(2014)Hermundstad, Brown, Bassett, Aminoff,
  Frithsen, Johnson, Tipper, Miller, Grafton, and
  Carlson}]{hermundstad2014structurally}
Hermundstad, A.~M., Brown, K.~S., Bassett, D.~S., Aminoff, E.~M., Frithsen, A.,
  Johnson, A., Tipper, C.~M., Miller, M.~B., Grafton, S.~T., Carlson, J.~M.,
  2014. Structurally-constrained relationships between cognitive states in the
  human brain. PLoS Comput Biol 10~(5), e1003591.

\bibitem[{Hilgetag and H{\"u}tt(2014)}]{hilgetag2014hierarchical}
Hilgetag, C.~C., H{\"u}tt, M.-T., 2014. Hierarchical modular brain connectivity
  is a stretch for criticality. Trends in cognitive sciences 18~(3), 114--115.

\bibitem[{Honey et~al.(2007)Honey, K{\"o}tter, Breakspear, and
  Sporns}]{honey2007network}
Honey, C.~J., K{\"o}tter, R., Breakspear, M., Sporns, O., 2007. Network
  structure of cerebral cortex shapes functional connectivity on multiple time
  scales. Proceedings of the National Academy of Sciences 104~(24),
  10240--10245.

\bibitem[{Hutchison et~al.(2013)Hutchison, Womelsdorf, Allen, Bandettini,
  Calhoun, Corbetta, Della~Penna, Duyn, Glover, Gonzalez-Castillo,
  et~al.}]{hutchison2013dynamic}
Hutchison, R.~M., Womelsdorf, T., Allen, E.~A., Bandettini, P.~A., Calhoun,
  V.~D., Corbetta, M., Della~Penna, S., Duyn, J.~H., Glover, G.~H.,
  Gonzalez-Castillo, J., et~al., 2013. Dynamic functional connectivity:
  promise, issues, and interpretations. Neuroimage 80, 360--378.

\bibitem[{Jones and Cercignani(2010)}]{jones2010twenty}
Jones, D.~K., Cercignani, M., 2010. Twenty-five pitfalls in the analysis of
  diffusion mri data. NMR in Biomedicine 23~(7), 803--820.

\bibitem[{Karrer et~al.(2008)Karrer, Levina, and Newman}]{karrer2008robustness}
Karrer, B., Levina, E., Newman, M.~E., 2008. Robustness of community structure
  in networks. Physical Review E 77~(4), 046119.

\bibitem[{Kinouchi and Copelli(2006)}]{kinouchi2006optimal}
Kinouchi, O., Copelli, M., 2006. Optimal dynamical range of excitable networks
  at criticality. Nature physics 2~(5), 348--351.

\bibitem[{Kivel{\"a} et~al.(2014)Kivel{\"a}, Arenas, Barthelemy, Gleeson,
  Moreno, and Porter}]{kivela2014multilayer}
Kivel{\"a}, M., Arenas, A., Barthelemy, M., Gleeson, J.~P., Moreno, Y., Porter,
  M.~A., 2014. Multilayer networks. Journal of complex networks 2~(3),
  203--271.

\bibitem[{Lambiotte et~al.(2014)Lambiotte, Delvenne, and
  Barahona}]{lambiotte2014random}
Lambiotte, R., Delvenne, J.-C., Barahona, M., 2014. Random walks, markov
  processes and the multiscale modular organization of complex networks. IEEE
  Transactions on Network Science and Engineering 1~(2), 76--90.

\bibitem[{Lancichinetti and Fortunato(2011)}]{lancichinetti2011}
Lancichinetti, A., Fortunato, S., 2011. Limits of modularity maximization in
  community detection. Phys Rev E Stat Nonlin Soft Matter Phys 84~(6 Pt 2),
  066122.

\bibitem[{Lohse et~al.(2014)Lohse, Bassett, Lim, and
  Carlson}]{lohse2014resolving}
Lohse, C., Bassett, D.~S., Lim, K.~O., Carlson, J.~M., 2014. Resolving
  anatomical and functional structure in human brain organization: identifying
  mesoscale organization in weighted network representations. PLoS Comput Biol
  10~(10), e1003712.

\bibitem[{Mantzaris et~al.(2013)Mantzaris, Bassett, Wymbs, Estrada, Porter,
  Mucha, Grafton, and Higham}]{mantzaris2013dynamics}
Mantzaris, A.~V., Bassett, D.~S., Wymbs, N.~F., Estrada, E., Porter, M.~A.,
  Mucha, P.~J., Grafton, S.~T., Higham, D.~J., 2013. Dynamics network
  centrality summarizes learning in the human brain. Journal of Complex
  Networks 1, 83--92.

\bibitem[{Mattar et~al.(2016)Mattar, Betzel, and Bassett}]{mattar2016flexible}
Mattar, M.~G., Betzel, R.~F., Bassett, D.~S., 2016. The flexible brain. Brain
  139~(Pt 8), 2110--2112.

\bibitem[{Mattar et~al.(2015)Mattar, Cole, Thompson-Schill, and
  Bassett}]{mattar2015functional}
Mattar, M.~G., Cole, M.~W., Thompson-Schill, S.~L., Bassett, D.~S., 2015. A
  functional cartography of cognitive systems. PLoS Comput Biol 11~(12),
  e1004533.

\bibitem[{Meunier et~al.(2009)Meunier, Lambiotte, Fornito, Ersche, and
  Bullmore}]{meunier2009}
Meunier, D., Lambiotte, R., Fornito, A., Ersche, K.~D., Bullmore, E.~T., 2009.
  Hierarchical modularity in human brain functional networks. Front Neuroinform
  3, 37.

\bibitem[{Moody and White(2003)}]{moody2003structural}
Moody, J., White, D.~R., 2003. Structural cohesion and embeddedness: A
  hierarchical concept of social groups. American Sociological Review,
  103--127.

\bibitem[{Mucha et~al.(2010)Mucha, Richardson, Macon, Porter, and
  Onnela}]{mucha2010}
Mucha, P.~J., Richardson, T., Macon, K., Porter, M.~A., Onnela, J.~P., 2010.
  Community structure in time-dependent, multiscale, and multiplex networks.
  Science 328~(5980), 876--878.

\bibitem[{Mueller et~al.(2013)Mueller, Wang, Fox, Yeo, Sepulcre, Sabuncu,
  Shafee, Lu, and Liu}]{mueller2013individual}
Mueller, S., Wang, D., Fox, M.~D., Yeo, B.~T., Sepulcre, J., Sabuncu, M.~R.,
  Shafee, R., Lu, J., Liu, H., 2013. Individual variability in functional
  connectivity architecture of the human brain. Neuron 77~(3), 586--595.

\bibitem[{Newman(2010)}]{newman2010networks}
Newman, M., 2010. Networks: an introduction. Oxford university press.

\bibitem[{Newman(2006)}]{newman2006modularity}
Newman, M.~E., 2006. Modularity and community structure in networks. Proc Natl
  Acad Sci U S A 103~(23), 8577--8582.

\bibitem[{Newman and Girvan(2004)}]{PhysRevE.69.026113}
Newman, M. E.~J., Girvan, M., Feb 2004. Finding and evaluating community
  structure in networks. Phys. Rev. E 69, 026113.
\newline\urlprefix\url{http://link.aps.org/doi/10.1103/PhysRevE.69.026113}

\bibitem[{Ozaktas(1992)}]{ozaktas1992paradigms}
Ozaktas, H.~M., 1992. Paradigms of connectivity for computer circuits and
  networks. Optical Engineering 31~(7), 1563--1567.

\bibitem[{Papadopoulos et~al.(2016)Papadopoulos, Puckett, Daniels, and
  Bassett}]{papadopoulos2016evolution}
Papadopoulos, L., Puckett, J.~G., Daniels, K.~E., Bassett, D.~S., 2016.
  Evolution of network architecture in a granular material under compression.
  Phys Rev E 94~(3-1), 032908.

\bibitem[{Park and Friston(2013)}]{park2013structural}
Park, H.-J., Friston, K., 2013. Structural and functional brain networks: from
  connections to cognition. Science 342~(6158), 1238411.

\bibitem[{Paul and Chen(2016)}]{paul2016null}
Paul, S., Chen, Y., 2016. Null models and modularity based community detection
  in multi-layer networks. arXiv preprint arXiv:1608.00623.

\bibitem[{Pons and Latapy(2005)}]{pons2005computing}
Pons, P., Latapy, M., 2005. Computing communities in large networks using
  random walks. In: International Symposium on Computer and Information
  Sciences. Springer, pp. 284--293.

\bibitem[{Porter et~al.(2009{\natexlab{a}})Porter, Onnela, and
  Mucha}]{porter2009}
Porter, M.~A., Onnela, J.-P., Mucha, P.~J., 2009{\natexlab{a}}. Communities in
  networks. Notices of the AMS 56~(9), 1082--1097.

\bibitem[{Porter et~al.(2009{\natexlab{b}})Porter, Onnela, and
  Mucha}]{porter2009communities}
Porter, M.~A., Onnela, J.-P., Mucha, P.~J., 2009{\natexlab{b}}. Communities in
  networks. Notices of the AMS 56~(9), 1082--1097.

\bibitem[{Power et~al.(2011)Power, Cohen, Nelson, Wig, Barnes, Church, Vogel,
  Laumann, Miezin, Schlaggar, and Petersen}]{power2011}
Power, J.~D., Cohen, A.~L., Nelson, S.~M., Wig, G.~S., Barnes, K.~A., Church,
  J.~A., Vogel, A.~C., Laumann, T.~O., Miezin, F.~M., Schlaggar, B.~L.,
  Petersen, S.~E., 2011. Functional network organization of the human brain.
  Neuron 72~(4), 665--678.

\bibitem[{Power et~al.(2014)Power, Mitra, Laumann, Snyder, Schlaggar, and
  Petersen}]{power2014methods}
Power, J.~D., Mitra, A., Laumann, T.~O., Snyder, A.~Z., Schlaggar, B.~L.,
  Petersen, S.~E., 2014. Methods to detect, characterize, and remove motion
  artifact in resting state {fMRI}. Neuroimage, 320--341.

\bibitem[{Preti et~al.(2016)Preti, Bolton, and Van De~Ville}]{preti2016dynamic}
Preti, M.~G., Bolton, T.~A., Van De~Ville, D., 2016. The dynamic functional
  connectome: {S}tate-of-the-art and perspectives. Neuroimage S1053--8119~(16),
  30788--30781.

\bibitem[{Pruim et~al.(2015)Pruim, Mennes, Buitelaar, and
  Beckmann}]{prium2015evaluation}
Pruim, R.~H., Mennes, M., Buitelaar, J.~K., Beckmann, C.~F., 2015. Evaluation
  of {ICA-AROMA} and alternative strategies for motion artifact removal in
  resting state {fMRI}. Neuroimage 112, 278--287.

\bibitem[{Rajan et~al.(2016)Rajan, Harvey, and Tank}]{rajan2016recurrent}
Rajan, K., Harvey, C.~D., Tank, D.~W., 2016. Recurrent network models of
  sequence generation and memory. Neuron 90~(1), 128--142.

\bibitem[{Reichardt and Bornholdt(2004)}]{reichardt2004detecting}
Reichardt, J., Bornholdt, S., 2004. Detecting fuzzy community structures in
  complex networks with a potts model. Physical Review Letters 93~(21), 218701.

\bibitem[{Reichardt and Bornholdt(2006)}]{reichardt2006statistical}
Reichardt, J., Bornholdt, S., 2006. Statistical mechanics of community
  detection. Physical Review E 74~(1), 016110.

\bibitem[{Ronhovde and Nussinov(2009)}]{ronhovde2009multiresolution}
Ronhovde, P., Nussinov, Z., 2009. Multiresolution community detection for
  megascale networks by information-based replica correlations. Physical Review
  E 80~(1), 016109.

\bibitem[{Rubinov and Sporns(2010)}]{rubinov2010complex}
Rubinov, M., Sporns, O., 2010. Complex network measures of brain connectivity:
  uses and interpretations. Neuroimage 52~(3), 1059--1069.

\bibitem[{Rykhlevskaia et~al.(2008)Rykhlevskaia, Gratton, and
  Fabiani}]{rykhlevskaia2008combining}
Rykhlevskaia, E., Gratton, G., Fabiani, M., 2008. Combining structural and
  functional neuroimaging data for studying brain connectivity: a review.
  Psychophysiology 45~(2), 173--187.

\bibitem[{Satterthwaite et~al.(2013)Satterthwaite, Elliott, Gerraty, Ruparel,
  Loughead, Calkins, Eickhoff, Hakonarson, Gur, Gur, and
  Wolf}]{satterthwaite2013improved}
Satterthwaite, T.~D., Elliott, M.~A., Gerraty, R.~T., Ruparel, K., Loughead,
  J., Calkins, M.~E., Eickhoff, S.~B., Hakonarson, H., Gur, R.~C., Gur, R.~E.,
  Wolf, D.~H., 2013. An improved framework for confound regression and
  filtering for control of motion artifact in the preprocessing of
  resting-state functional connectivity data. Neuroimage 64, 240--256.

\bibitem[{Schultz(1998)}]{schultz1998predictive}
Schultz, W., 1998. Predictive reward signal of dopamine neurons. Journal of
  neurophysiology 80~(1), 1--27.

\bibitem[{Schultz et~al.(1997)Schultz, Dayan, and Montague}]{schultz1997neural}
Schultz, W., Dayan, P., Montague, P.~R., 1997. A neural substrate of prediction
  and reward. Science 275~(5306), 1593--1599.

\bibitem[{Sherman and Guillery(2006)}]{sherman2006exploring}
Sherman, S., Guillery, R., 2006. Exploring the Thalamus and Its Role in
  Cortical Function. MIT Press.
\newline\urlprefix\url{https://books.google.com/books?id=xHyvQgAACAAJ}

\bibitem[{Siebenhuhner et~al.(2013)Siebenhuhner, Weiss, Coppola, Weinberger,
  and Bassett}]{siebenhuhner2013intra}
Siebenhuhner, F., Weiss, S.~A., Coppola, R., Weinberger, D.~R., Bassett, D.~S.,
  2013. Intra- and inter-frequency brain network structure in health and
  schizophrenia. PLoS One 8~(8), e72351.

\bibitem[{Simon(1991)}]{simon1991architecture}
Simon, H.~A., 1991. The architecture of complexity. In: Facets of systems
  science. Springer, pp. 457--476.

\bibitem[{Sizemore et~al.(2016)Sizemore, Giusti, Betzel, and
  Bassett}]{sizemore2016closures}
Sizemore, A., Giusti, C., Betzel, R.~F., Bassett, D.~S., 2016. Closures and
  cavities in the human connectome. arXiv preprint arXiv:1608.03520.

\bibitem[{Sporns(2010)}]{sporns2010networks}
Sporns, O., 2010. Networks of the Brain. MIT press.

\bibitem[{Sporns(2013)}]{sporns2013structure}
Sporns, O., 2013. Structure and function of complex brain networks. Dialogues
  Clin Neurosci 15~(3), 247--262.

\bibitem[{Sporns and Betzel(2016)}]{sporns2016modular}
Sporns, O., Betzel, R.~F., 2016. Modular brain networks. Annu Rev Psychol 67,
  613--640.

\bibitem[{Steinhaeuser et~al.(2012)Steinhaeuser, Ganguly, and
  Chawla}]{steinhaeuser2012multivariate}
Steinhaeuser, K., Ganguly, A.~R., Chawla, N.~V., 2012. Multivariate and
  multiscale dependence in the global climate system revealed through complex
  networks. Climate dynamics 39~(3-4), 889--895.

\bibitem[{S{\"u}mb{\"u}l et~al.(2014)S{\"u}mb{\"u}l, Song, McCulloch, Becker,
  Lin, Sanes, Masland, and Seung}]{sumbul2014genetic}
S{\"u}mb{\"u}l, U., Song, S., McCulloch, K., Becker, M., Lin, B., Sanes, J.~R.,
  Masland, R.~H., Seung, H.~S., 2014. A genetic and computational approach to
  structurally classify neuronal types. Nature communications 5.

\bibitem[{Swanson(2000)}]{swanson2000cerebral}
Swanson, L.~W., 2000. Cerebral hemisphere regulation of motivated behavior.
  Brain research 886~(1), 113--164.

\bibitem[{Traag et~al.(2011)Traag, Van~Dooren, and Nesterov}]{traag2011narrow}
Traag, V.~A., Van~Dooren, P., Nesterov, Y., 2011. Narrow scope for
  resolution-limit-free community detection. Phys Rev E Stat Nonlin Soft Matter
  Phys 84~(1 Pt 2), 016114.

\bibitem[{Tzourio-Mazoyer et~al.(2002)Tzourio-Mazoyer, Landeau, Papathanassiou,
  Crivello, Etard, Delcroix, Mazoyer, and Joliot}]{tzourio2002automated}
Tzourio-Mazoyer, N., Landeau, B., Papathanassiou, D., Crivello, F., Etard, O.,
  Delcroix, N., Mazoyer, B., Joliot, M., 2002. Automated anatomical labeling of
  activations in spm using a macroscopic anatomical parcellation of the mni mri
  single-subject brain. Neuroimage 15~(1), 273--289.

\bibitem[{Valverde et~al.(2015)Valverde, Ohse, Turalska, West, and
  Garcia-Ojalvo}]{valverde2015structural}
Valverde, S., Ohse, S., Turalska, M., West, B.~J., Garcia-Ojalvo, J., 2015.
  Structural determinants of criticality in biological networks. Frontiers in
  physiology 6.

\bibitem[{Vatansever et~al.(2015)Vatansever, Menon, Manktelow, Sahakian, and
  Stamatakis}]{vatansever2015default}
Vatansever, D., Menon, D.~K., Manktelow, A.~E., Sahakian, B.~J., Stamatakis,
  E.~A., 2015. Default mode network connectivity during task execution.
  Neuroimage 122, 96--104.

\bibitem[{Villegas et~al.(2014)Villegas, Moretti, and
  Mu{\~n}oz}]{villegas2014frustrated}
Villegas, P., Moretti, P., Mu{\~n}oz, M.~A., 2014. Frustrated hierarchical
  synchronization and emergent complexity in the human connectome network.
  Scientific reports 4.

\bibitem[{Vincent et~al.(2007)Vincent, Patel, Fox, Snyder, Baker, Van~Essen,
  Zempel, Snyder, Corbetta, and Raichle}]{vincent2007intrinsic}
Vincent, J., Patel, G., Fox, M., Snyder, A., Baker, J., Van~Essen, D., Zempel,
  J., Snyder, L., Corbetta, M., Raichle, M., 2007. Intrinsic functional
  architecture in the anaesthetized monkey brain. Nature 447~(7140), 83--86.

\bibitem[{Werner(2009)}]{werner2009fractals}
Werner, G., 2009. Fractals in the nervous system: conceptual implications for
  theoretical neuroscience. arXiv preprint arXiv:0910.2741.

\bibitem[{Womelsdorf et~al.(2014)Womelsdorf, Valiante, Sahin, Miller, and
  Tiesinga}]{womelsdorf2014dynamic}
Womelsdorf, T., Valiante, T.~A., Sahin, N.~T., Miller, K.~J., Tiesinga, P.,
  2014. Dynamic circuit motifs underlying rhythmic gain control, gating and
  integration. Nature neuroscience 17~(8), 1031--1039.

\bibitem[{Wymbs et~al.(2012)Wymbs, Bassett, Mucha, Porter, and
  Grafton}]{wymbs2012differential}
Wymbs, N.~F., Bassett, D.~S., Mucha, P.~J., Porter, M.~A., Grafton, S.~T.,
  2012. Differential recruitment of the sensorimotor putamen and frontoparietal
  cortex during motor chunking in humans. Neuron 74~(5), 936--946.

\bibitem[{Yaesoubi et~al.(2015)Yaesoubi, Allen, Miller, and
  Calhoun}]{yaesoubi2015dynamic}
Yaesoubi, M., Allen, E.~A., Miller, R.~L., Calhoun, V.~D., 2015. Dynamic
  coherence analysis of resting fmri data to jointly capture state-based phase,
  frequency, and time-domain information. NeuroImage 120, 133--142.

\bibitem[{Yeh and Tseng(2011)}]{yeh2011ntu}
Yeh, F.-C., Tseng, W.-Y.~I., 2011. Ntu-90: a high angular resolution brain
  atlas constructed by q-space diffeomorphic reconstruction. Neuroimage 58~(1),
  91--99.

\bibitem[{Yeh et~al.(2013)Yeh, Verstynen, Wang, Fern{\'a}ndez-Miranda, and
  Tseng}]{yeh2013deterministic}
Yeh, F.-C., Verstynen, T.~D., Wang, Y., Fern{\'a}ndez-Miranda, J.~C., Tseng,
  W.-Y.~I., 2013. Deterministic diffusion fiber tracking improved by
  quantitative anisotropy. PloS one 8~(11), e80713.

\bibitem[{Yeo et~al.(2011)Yeo, Krienen, Sepulcre, Sabuncu, Lashkari,
  Hollinshead, Roffman, Smoller, Zollei, Polimeni, Fischl, Liu, and
  Buckner}]{yeo2011}
Yeo, B.~T., Krienen, F.~M., Sepulcre, J., Sabuncu, M.~R., Lashkari, D.,
  Hollinshead, M., Roffman, J.~L., Smoller, J.~W., Zollei, L., Polimeni, J.~R.,
  Fischl, B., Liu, H., Buckner, R.~L., 2011. The organization of the human
  cerebral cortex estimated by intrinsic functional connectivity. J
  Neurophysiol 106~(3), 1125--1165.

\bibitem[{Zhang et~al.(2016)Zhang, Telesford, Giusti, Lim, and
  Bassett}]{zhang2016choosing}
Zhang, Z., Telesford, Q.~K., Giusti, C., Lim, K.~O., Bassett, D.~S., 2016.
  Choosing wavelet methods, filters, and lengths for functional brain network
  construction. PLoS One 11~(6), e0157243.

\bibitem[{Zhang et~al.(2007)Zhang, Zhou, and Zou}]{zhang2007self}
Zhang, Z.-Z., Zhou, S.-G., Zou, T., 2007. Self-similarity, small-world,
  scale-free scaling, disassortativity, and robustness in hierarchical
  lattices. The European Physical Journal B 56~(3), 259--271.

\bibitem[{Zhou et~al.(2006)Zhou, Zemanov\'a, Zamora, Hilgetag, and
  Kurths}]{PhysRevLett.97.238103}
Zhou, C., Zemanov\'a, L., Zamora, G., Hilgetag, C.~C., Kurths, J., Dec 2006.
  Hierarchical organization unveiled by functional connectivity in complex
  brain networks. Phys. Rev. Lett. 97, 238103.
\newline\urlprefix\url{http://link.aps.org/doi/10.1103/PhysRevLett.97.238103}

\bibitem[{Zilles and Amunts(2013)}]{zilles2013individual}
Zilles, K., Amunts, K., 2013. Individual variability is not noise. Trends in
  cognitive sciences 17~(4), 153--155.

\end{thebibliography}





\end{document}